\documentclass[%
pra, superscriptaddress, twocolumn,
amsmath, amssymb, aps, pra, 
floatfix, ]{revtex4-1}
\usepackage{nicefrac}
\usepackage{graphicx}
\usepackage{bm}
\usepackage{todonotes}
\usepackage[protrusion=true,expansion=true,tracking=true]{microtype}
\usepackage{hyperref}
\usepackage[capitalize]{cleveref}
\usepackage{dsfont}
\usepackage{braket}
\usepackage{xcolor}
\usepackage{cleveref}
\usepackage[caption=false]{subfig}

\usepackage{changes}
\usepackage{cancel}

\crefname{section}{Sec.}{Sec.}

\renewcommand{\vec}[1]{{\bm{#1}}}

\newcommand{\im}{\mathrm{Im}}

\newcommand{\tr}{\mathrm{Tr}}
\newcommand{\del}[1]{\partial_{#1}}
\newcommand{\clcm}{C_\mathrm{LCM}}

\newcommand{\mycaption}[2]{\caption[#1]{\emph{#1} #2}}
\newcommand{\hc}{\ensuremath{\textrm{h.c.}}}

\newcommand{\unit}{\ensuremath{\mathds{1}}}

\begin{document}
    \title{Effect of disorder on topological charge pumping in the Rice-Mele model}
\author{A.~L.~C.~Hayward}
\affiliation{Institute for Theoretical Physics, Georg-August-Universit\"at G\"ottingen,
Friedrich-Hund-Platz 1, 37077 G\"ottingen, Germany}
\affiliation{Arnold Sommerfeld Center for Theoretical Physics, Ludwig-Maximilians-Universit\"at M\"unchen, 80333 M\"unchen, Germany}
\author{E. Bertok}
\affiliation{Institute for Theoretical Physics, Georg-August-Universit\"at G\"ottingen,
Friedrich-Hund-Platz 1, 37077 G\"ottingen, Germany}
\author{U. Schneider}
\affiliation{Cavendish Laboratory, University of Cambridge, J.J. Thomson Avenue, Cambridge CB3 0HE, United Kingdom}
\author{F. Heidrich-Meisner}
\affiliation{Institute for Theoretical Physics, Georg-August-Universit\"at G\"ottingen,
Friedrich-Hund-Platz 1, 37077 G\"ottingen, Germany}

\date{\today}
\begin{abstract}
Recent experiments with ultracold quantum gases have successfully realized integer-quantized topological charge pumping in optical lattices.
Motivated by this progress, we study the effects of static disorder on topological Thouless charge pumping. We focus on the half-filled Rice-Mele model of free spinless fermions and consider random diagonal disorder.
We use both static  and time-dependent simulations to characterize the charge pump. 
The set of measures that we compute in the instantaneous basis include the polarization, the entanglement spectrum, and the space-integrated local
Chern marker. As a first main result, we conclude that the space-integrated local Chern marker is  best suited for a quantitative determination of  topological transitions in a disordered system.
In the time-dependent simulations, we use the time-integrated current to obtain the pumped charge in slowly periodically driven systems. 
As a second main result, we observe and characterize a disorder-driven breakdown of the quantized charge pump. There is  an excellent agreement between the static and the time-dependent ways of computing the pumped charge. The topological transition occurs well in the regime where all states are localized on the given system sizes. This observation is consistent with previous studies  and the topological transition  is therefore not tied to a delocalization-localization transition of Hamiltonian eigenstates. For individual disorder realizations,
the breakdown of the quantized pumping occurs for parameters  where the  spectral bulk gap inherited from the band gap of the clean system closes, leading to a globally gapless spectrum.
As a third main result and with respect to the analysis of finite-size systems, we show that the disorder average of the bulk gap severely overestimates the stability of quantized pumping.
A much better estimate is the typical value of the distribution of energy gaps, also called mode of the distribution.
We discuss our results in the context of recent quantum-gas experiments that realized charge pumps.
\end{abstract}

\maketitle
\section{Introduction}
 
The physics of topological charge pumps \cite{thouless1983quantization,niu1984quantised} has attracted intensified interest due to 
recent quantum-gas experiments realizing charge pumps with fermions \cite{Nakajima2016} or strongly
interacting bosons \cite{Lohse2016}, and a spin pump \cite{Schweizer2016}. 
These experiments observe quantized pumping over a number of cycles
despite the fact that finite particle numbers and inhomogeneous systems
were considered. 
Current theoretical efforts investigate the stability of charge pumps in optical lattices
in genuine many-body systems \cite{Berg2011,Rossini2013,Ke2017,Kuno2017,Nakagawa2018,Hayward2018,Qin2018,Mei2019,Stenzel2019,Lin2020,Greschner2020,Lin2020a, Marks2021},
in noninteracting systems with disorder  \cite{qin_quantum_2016,Imura2018,Wang2019,Ippoliti2020,Hu2020,Marra2020}, with dissipation \cite{Arceci2020}, or as a proximity effect \cite{Waver2020}. Moreover, there are recent conceptual developments as well, for instance, concerning the introduction of the boundary charge \cite{Lin2020a,Pletyukhov2020,Pletyukhov2020a,Weber2021}.

A common notion is that topological properties  remain protected against small amounts of disorder \cite{niu1984quantised,Qi2011,Hasan2010,Chiu2016,Cooper2019}.
Considering a  topological insulator, one expects the integer quantization of 
a topological invariant to remain robust as long as the bulk gap stays intact. 
Therefore, a disorder drawn from a bounded disorder distribution (e.g., a box distribution), whose width is significantly smaller than the many-body gap, does not lead to a breakdown of topological quantization (see, e.g., Ref.~\cite{Qi2011}).
Remarkably, disorder can also induce topological properties into a system, leading to the so-called topological Anderson insulator \cite{Li2009}, which has been investigated experimentally \cite{Meier2018,Nakajima2020}.
Different from the scenario studied here, Titum {\it et al.} \cite{Titum2016} introduced the case of an anomalous Floquet-Anderson insulator that is characterized by a winding number rather than a 
Chern number and possess fully localized Floquet eigenstates. 

Another interesting direction concerns topology in quasicrystals \cite{Kraus2012}. Two recent experiments (using ultracold atoms \cite{Nakajima2020} and a photonic system \cite{Cerjan2020}) have investigated the stability of a topological charge pump with noninteracting particles against disorder.
The fermionic quantum-gas experiment \cite{Nakajima2020} has realized a quasiperiodic disorder potential that is akin to the Aubry-André model \cite{aubry1980analyticity}. Both disorder-induced quantized pumping as well as the breakdown of topology in a sufficiently strong disorder potential has been observed.

Note that Aubry-Andr\'{e}-like  systems have been realized in several quantum-gas experiments \cite{Schreiber2015,Rispoli2019, Bordia2017,Roati2008}, while more general
forms of quasicrystals are at the heart of recent experimental efforts  \cite{Dareau2017,Viebahn2019,Rajagopal2019,Xiao2020}.
A number of theoretical studies has addressed the stability of topological properties against this type
of quasi-disorder, including, e.g., the SSH model  \cite{liu_topological_2018}.
The effect of disorder on an SSH model has also been investigated in a recent quantum-gas experiment  \cite{Meier2018}.
 
\begin{figure}[t!]
   \centering
        \includegraphics{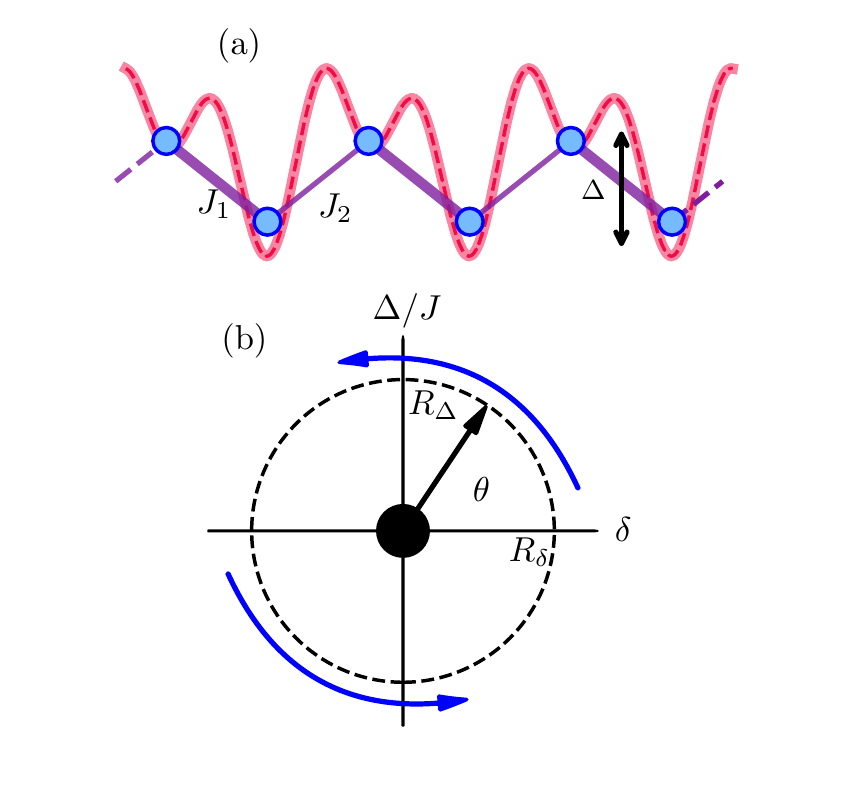}
        \mycaption{Rice-Mele model.}{ (a) A bichromatic potential
        with a ratio of two between the two optical wavelenghts  can, in the
      deep-lattice limit, be described by a  lattice model with alternating tunneling strength, $J_{1,2}$, and a staggered potential $\Delta$. (b) We consider a charge
pump that traces a path through this parameter space which encircles the central
degeneracy. The cycle is parameterized by the angle $\theta$, $R_\delta$, and $R_\Delta$.}
   \label{fig:model}
\end{figure}

Motivated by these experimental developments, we theoretically investigate the stability of topological properties of charge pumps against disorder in systems of noninteracting fermions.
 We concentrate on the half-filled Rice-Mele model with periodic boundary conditions   (see the sketch in Fig.~\ref{fig:model}) and introduce random disorder in the on-site potentials. 
The problem of charge pumping in the presence of disorder has previously been addressed in a number of related studies \cite{qin_quantum_2016,Nakajima2020,Wauters2019}.
In Ref.~\cite{qin_quantum_2016}, a different model has been investigated using open boundary conditions, with an emphasis on the pumping of charge
between edge states. As a result, they report  the existence of a disorder-driven topological transition into a trivial state.
A very appealing perspective on a disordered charge pump, modeled by the Rice-Mele model,  has been put forward in Ref.~\cite{Wauters2019}: While the Hamiltonian eigenstates are fully localized in one dimension with random onsite disorder for any disorder strength \cite{Abrahams1979}, there is a delocalization-localization transition in the Floquet eigenstates at the point where quantized pumping breaks down.
Our work extends the analysis and theoretical understanding in several ways.
First and complementary to the approach of \cite{qin_quantum_2016}, we use periodic boundary conditions that circumvent  the complication of identifying edge states. Their unambiguous identification  is complicated in the presence of randomness due to possible Anderson-localization of all single-particle states. 
Instead, we characterize the charge pump by four bulk quantities: The polarization, the entanglement spectrum, the local Chern marker and its integral as a measure for the bulk Chern number, and, finally, the integrated time-dependent current.  
The first three quantities are studied in the instantaneous eigenbasis, and the integrated current is extracted from time-dependent simulations with finite-period pump cycles.

We give a brief account of our main results.
The polarization is the most established object for the description of adiabatic transport  \cite{Resta1994}. We argue that 
 the representation of the polarization as an  angular variable  simplifies its analysis. 
 The entanglement spectrum 
contains information about quantized charge pumping via the spectral flow as has been  discussed in \cite{Hayward2018}. Here,
we show that the quantized winding is preserved in the presence of weak disorder, while symmetries protecting the topology of the Su-Schrieffer-Heeger (SSH) model \cite{Su1979} that is visited twice in a clean
system are broken.

Generally, in an inhomogeneous system, one cannot rely on the classification and expressions for topological invariants of lattice-translational invariant systems and hence a need for
real-space version arises, as was emphasized in \cite{Altland2015}. The usual path is to work with many-body wave functions and twisted boundary conditions, just as in the
description of the integer quantum Hall effect. The local Chern marker provides an alternative tool for the calculation of an invariant in inhomogeneous systems.
The local Chern marker
has originally been introduced to capture aspects of topology in spatially inhomogeneous and/or finite systems  \cite{Bianco2011} 
such as they arise in the presence of a trapping potential or interfaces (see, e.g., \cite{Irsigler2019}). Here, we use the local Chern marker
to obtain the global Chern number of our disordered systems by integration over the spatial coordinate. For this purpose, we generalize the expression for the local Chern marker to 
a time-periodic situation in the adiabatic limit. For the critical disorder strength, we observe an excellent agreement between the integrated local Chern marker and the pumped charge obtained from sufficiently slow time-dependent simulations. We stress that our work focuses on finite systems as they are realized in ultracold quantum gases. 
Recently, another form of the local Chern marker has been studied for odd dimensions by using an effective adiabatic propagator \cite{sykes2020}.

The analysis of the spatial dependence of the local Chern marker and its fluctuations around the bulk Chern number provides a useful quantitative measure for the topological transition.
We argue that the integrated local Chern marker and the integrated current are the best suited to identify the transition point at which quantized charge pumping breaks down. In particular,  the pumped charge, which is equivalent to the time-integrated current,  can readily be accessed in cold-atom experiments \cite{Lohse2016,Nakajima2016,Nakajima2020}. 

Since most quantum-gas experiments measure an average over many one-dimensional realizations with varying disorder potentials, it is important to carefully
analyze the full distribution of the relevant measures. We show that the disorder-average of the bulk gap overestimates the critical disorder strength, while the mode,
the most likely value, agrees very well with the point at which the disorder-averaged measures for topological charge pumping start to deviate from integer values. 
This insight may be relevant for the analysis of experimental data \cite{Nakajima2020} as well as of numerical studies of finite systems \cite{Wauters2019}. 

The plan of this exposition is the following. In Sec.~\ref{sec:model}, we introduce the Rice-Mele model and the types
of disorder studied in this work. Section~\ref{sec:methods} discusses the set of measures used to characterize the quantized charge pumping. 
Our results are presented in Sec.~\ref{sec:random}, where we discuss static measures and
the results of time-dependent simulations. We conclude with a summary and a discussion in Sec.~\ref{sec:summary}. In the Appendix, we discuss results from an additional parameter set.

\section{Model}
\label{sec:model}

The Rice-Mele model \cite{Rice1982} describes a one-dimensional
lattice system of spinless fermions with alternating hopping-matrix elements and a staggered on-site
potential, illustrated schematically in Fig.~\ref{fig:model}(a). It  can be written as:
\begin{equation}
  \label{eq:rm_model}
  H  =  \sum_{j=0}^{L-1} \left[ -J_j \hat c_j^\dagger \hat c_{j+1} + \hc
    +  V_j \hat n_j \right].
\end{equation}
Here, $j$ is the site index, $J_j = J(1+(-1)^j\delta)$ is an alternating tunneling rate [with $J_j=J_1(J_2)$ for $j$ even(odd)], and $V_j = \left (\frac{(-1)^j \Delta}{2}\right )$ is the onsite staggered potential.
For simplicity, we parametrize the alternating tunneling with the dimensionless dimerization
parameter $\delta \in [-1,1]$ and with $J= (J_1 + J_2)/2$ setting the energy scale. The  strength of the staggered potential is given by $\Delta$.
The operator $\hat c_j^\dagger$
creates a fermion on site $j$ and $\hat n_j = \hat c_j^\dagger \hat c_j$. We use a smooth pumping scheme of the form 
\begin{equation}
(\Delta/J,\delta) = (R_\Delta \sin \theta,  R_\delta \cos \theta)\,. \label{eq:cycle}
\end{equation}
Note that at $\theta = 0(\pi)$, the SSH model is realized in its topological(trivial) phase.

We  consider uniform, bounded diagonal disorder. 
An instance of random diagonal disorder results from a  modification of the onsite potential:
\begin{equation}
V_j \rightarrow V_j + \epsilon_j.
\end{equation}
Here, $\epsilon_j$ is an energy drawn from the uniform distribution $\epsilon_j \in \frac{w}{2}\left[-1,1\right] $, where $w$ is the disorder strength.

\section{Methods and Observables}
\label{sec:methods}

\subsection{Polarization}

 To characterize the topological properties of a charge
pump, one can consider the evolution of the many-body polarization $P(t)$
over the course of the adiabatic driving of a time-dependent Hamiltonian \cite{Resta1994}. 
The integral over the time derivative of $P$ yields the pumped charge:
\begin{equation}
  \label{eq:int-current}
  \Delta Q  =  \int_0^{\Delta t}dt\,  \partial_tP(t).
\end{equation}
For a system with translational invariance, the polarization is usually formulated in terms of the cell-periodic part of the momentum eigenstates in the unit cell, here denoted by $\ket{u(k)}$, leading to
\begin{equation}
P = \frac{i}{2\pi}\int dk \braket{u(k)| \partial_k u(k)}  \,. \end{equation}
In the case of a disordered system, the quasimomentum  $k$ is no longer a good quantum number and the topological invariants must be constructed in the position basis. 

For a finite system with periodic boundary conditions, we use the exponentiated position operator, which is, following Resta \cite{Resta1998}, given by 
  $\hat{X}^e = \exp\left(i\frac{2\pi}{L} \hat{X}\right) $. $\hat X = \sum_{j=1}^{L} j \hat c_j^\dagger \hat c_j$ is the total position operator. Then, we obtain:
\begin{equation}
  \label{eq:pol-MB1}
  P(t)=\frac{Qa}{2\pi}\operatorname{Im}\ln\left\langle\Psi(t) \middle|
  \hat{X}^e  \middle| \Psi(t)\right\rangle\quad(\bmod \,Qa),
\end{equation}
where $\left| \Psi(t)\right\rangle$ is a many-body wavefunction.
Importantly, $P$ is only defined modulo $Qa$, which has units of a dipole moment, while  $a$ is the lattice spacing set to unity and $Q$ symbolizes the charge.  This
expression for the many-body polarization reduces to the usual form for
 noninteracting fermions for a filled band
 \cite{Resta1992, King1993,Resta1994, watanabe_inequivalent_2018}.

For a many-body state, which is a Slater determinant composed of single-particle states $\psi_i$, the polarization can be written as 
\begin{equation}
  P(t)=\frac{Q a}{2\pi} \im \ln \det \Psi^\dagger e^{\frac{-i2\pi}{L} \hat X} \Psi. 
  \label{eq:pol-MB2}
\end{equation}
Here, $\Psi_{ij} = \psi_i(j)$ is different from  the many-body wavefunction in Eq.~(\ref{eq:pol-MB1}), 
and $\ket{\psi_i} $  are the occupied single-particle states.
The total transported charge $\Delta Q$ is related to the polarization via
Eq.~\eqref{eq:int-current}.

For a charge pump, we see
that the quantization of transported charge corresponds to the winding of the
polarization. 
This winding is necessarily quantized, which seems like a contradiction, i.e., it
seems to preclude  non-quantized pumping. This can be reconciled by realizing that, for
gapless states, the polarization becomes non-analytic \cite{Imura2018}.

\subsection{Local Chern Marker}
\label{sec:LCM_method}

The local Chern marker \cite{Bianco2011} (LCM) is a local observable that can capture some aspects of
topology from 'local' (real-space) properties of a system. In the present work, it is mainly used as a tool for computing the Chern number of inhomogeneous systems with periodic boundary conditions. 
Moreover, we will analyze the information contained in the spatial fluctuations of the local Chern marker $C(j)$. An alternative to the local Chern marker for the description of spatially inhomogeneous systems  has recently been used in \cite{Nakajima2020} using twisted boundary conditions.

We start from the expression for the Chern number of a spatially two-dimensional translational invariant  system with a quasimomentum
$(k_x,k_y)$ 
\begin{eqnarray}
  C &=& -\frac{1}{\pi}\im\int dk_x \int dk_y   \sum_{p\in \mathcal{B}} \sum_{q\notin \mathcal{B}}   \\
     && \braket{p,k_x,k_y| \partial_{k_x} |q,k_x,k_y}\braket{q,k_x,k_y| \partial_{k_y} |p,k_x,k_y}, \nonumber
\end{eqnarray}
where $\mathcal{B}$ is a filled band, or set of occupied states. 
To arrive at a real-space representation, we need to identify the projection operators onto the occupied and unoccupied single-particle subspaces. Inserting a resolution of the identity by summing over an additional set of occupied single-particle states $\ket{n,k_x,k_y}$ yields (omitting $k_x$ and $k_y$ quantum numbers in the states $\ket{n}$ to avoid cluttering the formula):
\begin{eqnarray}
  C &= -\frac{1}{\pi}\im\int dk_x \int dk_y   \sum_{n\in \mathcal{B}} \sum_{p\in \mathcal{B}} \sum_{p\notin \mathcal{B}} \nonumber  \\
     & \braket{p|n} \braket{n| \partial_{k_x} |q}\braket{q| \partial_{k_y} |p}.
\end{eqnarray}
By replacing the $k_x$ integral with a sum  over states, one can rewrite this as:
\begin{eqnarray}
  C &= -\frac{2}{L}\im\sum_{k_x} \int dk_y   \sum_{n} \sum_{p\in \mathcal{B}} \sum_{p\notin \mathcal{B}} \nonumber  \\
     & \braket{p|n} \braket{n| \partial_{k_x} |q}\braket{q| \partial_{k_y} |p}\\
  &= -\frac{2}{L}\;  \int dk_y \,\im \tr \left[ \hat{P} {\partial_{k_x} } \hat{Q} {\partial_{k_y} } \hat{P} \right],
\label{eq:lcm_start}
\end{eqnarray}
where we use the cyclic property of the trace, and the trace is here expressed via  
\begin{eqnarray}
  \tr\; \cdot =\sum_{n} \sum_{k_x} \braket{n,k_x,k_y|\cdot|n,k_x,k_y}
\end{eqnarray} and
\begin{eqnarray}
  &\partial_{k_y} \hat P = \sum_{p\in \mathcal{B}} \left(\partial_{k_y}\ket{p,k_x,k_y}\right)\bra{p,k_x,k_y} \nonumber\\
  & + \ket{p,k_x,k_y} \left(\partial_{k_y}\bra{p,k_x,k_y}\right).
\end{eqnarray}
Here, $\hat{P}$ and $\hat{Q} = \unit - \hat{P}$ are projection operators onto the occupied and unoccupied states, respectively. The local Chern marker $C(\vec r)$ is found by inserting a complete set of basis states $\ket{\mathbf{r}}$ in the position basis:
\begin{equation}
\label{bulk_lcm}
  C_{\rm LCM} = -\frac{2}{L}\; \im \sum_\mathbf{r} C(\mathbf{r}) = \frac{2}{L}\;\im \sum_\mathbf{r} \bra{\mathbf{r}} \hat{P} {\partial_{k_x} } \hat{Q} {\partial_{k_y} } \hat{P} \ket{\mathbf{r}}.
\end{equation}
The result is the bulk Chern number $C_{\rm LCM}$, where the subindex LCM indicates that it is computed from the local Chern marker.
Equation~\eqref{bulk_lcm} is the usual expression for the local Chern marker (by omitting the sum over $\mathbf{r}$) for a spatially inhomogeneous system in two spatial dimensions.

Using the correspondence for an adiabatic periodic system with pumping frequency $\Omega$ and a synthetic dimension, under which $\Omega t = \theta \leftrightarrow k_y$, we can find an analogous expression for a one-dimensional charge pump, starting from Eq.~\eqref{eq:lcm_start}. However, as there is no analogous real-space operator for the time dimension, the derivative and integral with respect to time has to be computed explicitly while the trace over the site index $j$ is omitted in  Eq.~\eqref{eq:lcm_start}:
\begin{equation}
  C(j) = - \frac{2}{L}\;\im \int\limits_0^{2\pi} d \theta \braket{j|\hat P \partial_{k} \hat Q \partial_\theta \hat P|j}.
  \label{eq:lcmprelim}
\end{equation}
For the $k$ derivative,
we obtain, using the product rule

  \begin{equation}
    \hat P(\partial_k \hat Q) = \partial_k (\hat P\hat Q) - (\partial_k \hat P)\hat Q = - (\partial_k \hat P)\hat Q,
    \label{eq:leibnitz}
    \end{equation}

since $\hat P\hat Q=0$.
For a finite system, the minimum quasi-momentum difference is  $dk = 2\pi/L$. Hence, we may replace the $k$ derivative in Eq.~\eqref{eq:leibnitz} by the finite-difference form:
  \begin{align}
    (\partial_k \hat P)\hat Q &=\frac{\hat P (k+dk)-\hat P(k)}{dk}\hat Q(k)=\frac{\hat P (k+dk)}{dk}\hat Q(k) \nonumber\\
    &= \frac{L}{2\pi} \hat{X}^{e\dagger} \hat P(k) \hat{X}^{e} \hat Q(k),
    \label{eq:leibnitz2}
  \end{align}

where in the first line,  $\hat P \hat Q = 0$ was used. In the second line, the exponential position operator $\hat{X}^e = \exp\left(i\frac{2\pi}{L} \hat{X}\right)$ acts as the generator of momentum translations of $dk=2\pi/L$, which is applied from both sides to the operator $\hat P(k)$.
Inserting Eqs.~\eqref{eq:leibnitz} and \eqref{eq:leibnitz2} into Eq.~\eqref{eq:lcmprelim}, we thus arrive at the LCM for a charge pump with periodic boundary conditions:
\begin{align}
  C(j)=\im \frac{1}{\pi}\int\limits_0^{2\pi}d\theta \braket{j|\hat X^{e\dagger}\hat P(\theta) \hat X^{e}\hat Q(\theta) (\del{\theta} \hat P(\theta) )|j}.
  \label{equ:time-lcm}
\end{align}

Translation-invariant systems (with periodic boundary conditions) will have a constant LCM. The introduction of disorder breaks the translational symmetry, leading to a fluctuating, position-dependent  LCM. However, so long as the system has a band gap, the sum $C_\mathrm{LCM} = \sum_j C(j)$ of the LCM is expected to be quantized. 

In the case of a gapless system, the kernel of Eq.~\eqref{equ:time-lcm} is discontinuous. This leads to non-quantized values for $\clcm$  when the system is gapless, indicating a breakdown of quantized pumping.

To compute the local Chern marker, we use a discretized form with periodic boundary conditions. Using that $\hat Q \partial_\theta \hat P = \hat Q (\partial_\theta \hat P)\hat{P} = \lim_{d\theta\rightarrow 0} \hat Q(\theta)\hat P(\theta+d\theta) \hat P(\theta)/d\theta$, we arrive at
\begin{equation}
  C(j) = \frac{1}{\pi}\sum_{n=0}^{N_\theta-1} \im \bra{j} \hat{X}^{e\dagger} \hat{P}(\theta_n) \hat{X}^e \hat{Q}(\theta_n) \hat{P}(\theta_{n+1}) \hat{P}(\theta_n)\ket{j}.
  \label{eq:lcmnumerical}
\end{equation}
Here, $\theta_n = 2\pi n /N_\theta$, and $N_\theta$ is the number of points in the discretization of $\theta$ with a step size $d\theta = \theta_{n+1}-\theta_{n}$.

Our numerical analysis shows that in the topological phase, this converges to a fixed value for $C(j)$ as $d\theta$ is decreased. The projector $\hat{P}(\theta$) is not necessarily continuous. 
However, the integral over this function is well-defined and converges with decreasing $d\theta$. 
The sum $\clcm:=\sum_j C(j)$ yields the Chern number for a translationally invariant system as our discussion shows.

\begin{figure}[tb]
  \centering \includegraphics{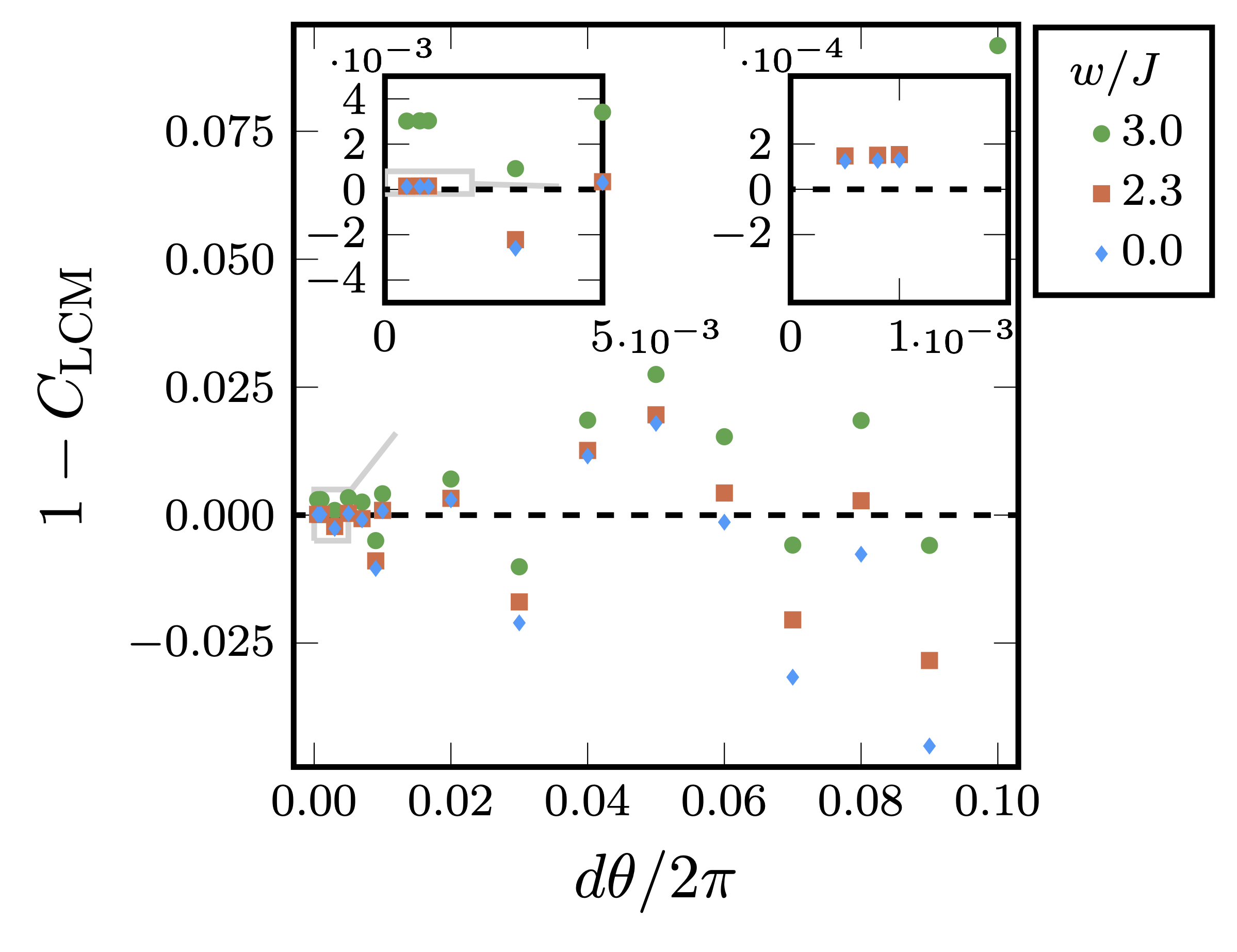}
  \mycaption{Convergence of the integrated local Chern marker.}
    {Main panel: LCM calculated for a single disorder realization as a function of discretization $d\theta/2\pi$ in the sum in Eq.~(\ref{eq:lcmnumerical}) for $L=400$ and for $w/J = 0,2.3,3$ (colors). The dashed line indicates a perfect quantization of the Chern number of 1. Insets: Left: Zoom into the main panel for small discretizations. Right: Further zoom, showing the region of convergence beyond $d\theta/2\pi=1\cdot10^{-3}$.
   } \label{fig:lcmconvergence}
  \end{figure}

In the simulations, we use $d\theta/2\pi=10^{-3}$ which gives a Chern number which is quantized with an accuracy of $1-C_\mathrm{LCM}=\pm10^{-4}$ at $w=0$ and for individual realizations for  $w/J\lesssim 2$, which is shown in Fig.~\ref{fig:lcmconvergence}.
In the trivial phase at $w/J>3$, a sizable dependence on $d\theta$ of the LCM can persist even up to $d\theta/2\pi=10^{-5}$ for individual realizations (data not shown).
This can lead to a smaller accuracy of $10^{-1}$.
Therefore, the quantitative results for the LCM in the trivial phase are less reliable for single disorder realizations. However, due to the disorder average, the mean of the integrated LCM $\overline\clcm$ is two orders of magnitude more reliable when compared to single realizations.

 We will apply this approach to inhomogeneous systems 
in this work and the results and comparison with other measures will lead to consistent results for integer quantization of $\clcm$ in the topological phase.

\subsection{Entanglement spectrum}

The entanglement spectrum is given by the eigenvalues of the entanglement Hamiltonian $ \hat H_E$ and can be defined with the reduced density matrix $\hat \rho_A$ of a spatial bipartition of the system into two halves ($A$ and $B$)  \cite{Li2008}:
\begin{equation}
	\frac{1}{Z}\hat\rho_A = e^{-\hat H_E} \,.
\end{equation}

$\hat H_E$ is a dimensionless free-fermion operator that  is strictly positive due to the normalization requirement of the reduced density matrix $\hat \rho_A$.
The entanglement eigenvalues (EEVs) are the eigenvalues of $\hat H_E$.
The EEVs $ -\mbox{ln} \Lambda^2_\mu$ are related to the Schmidt values $\Lambda_\mu$ that are found in a Schmidt decomposition of the two halves of the system. 
We can compute these values of the free-fermion Hamiltonian directly from the reduced density matrix, which can be obtained from the single-particle correlation matrix $C_{ij}=\braket{\hat c_i^\dagger \hat c_j}$ computed in the many-body ground state,  as described by Peschel \cite{Peschel2009}. 
Writing $\hat H_E$ in its single-particle eigenbasis, $\hat H_E = \sum_{\alpha=1}^L \epsilon_\alpha \hat d_\alpha^\dagger \hat d_\alpha$, the partition function is $Z=\prod_{\alpha=1}^L \frac{1}{1+e^{-\epsilon_\alpha}}$. 
The eigenvalues $\Lambda_\mu^2$ in the full many-body space of the reduced density matrix $\hat \rho_A$ result from fixing a set of  occupations $\vec{n} = \{n_\alpha\}$. Every choice for these occupations corresponds to a  many-body state $\ket{\mu}$ indexed by $\mu$. They are given by 
\begin{equation}
  -\ln(\Lambda_\mu^2)=\sum_{\alpha=1}^{L}\left[ -\ln(v_\alpha)n_\alpha - \ln(1-v_\alpha)(1-n_\alpha)\right],
  \label{eq:fullrho}
\end{equation}
where  $v_\alpha$ are the eigenvalues of the single-particle correlation matrix $\braket{\hat c_i^\dagger \hat c_j}$ when $i,j$ are restricted to one half of the system. 
As the states are all localized for finite disorder, the entanglement across one cut in the system is independent of the boundary conditions for a sufficiently large system. The same is true for zero disorder due to translational invariance.

\begin{figure}[tb]
  \centering \includegraphics{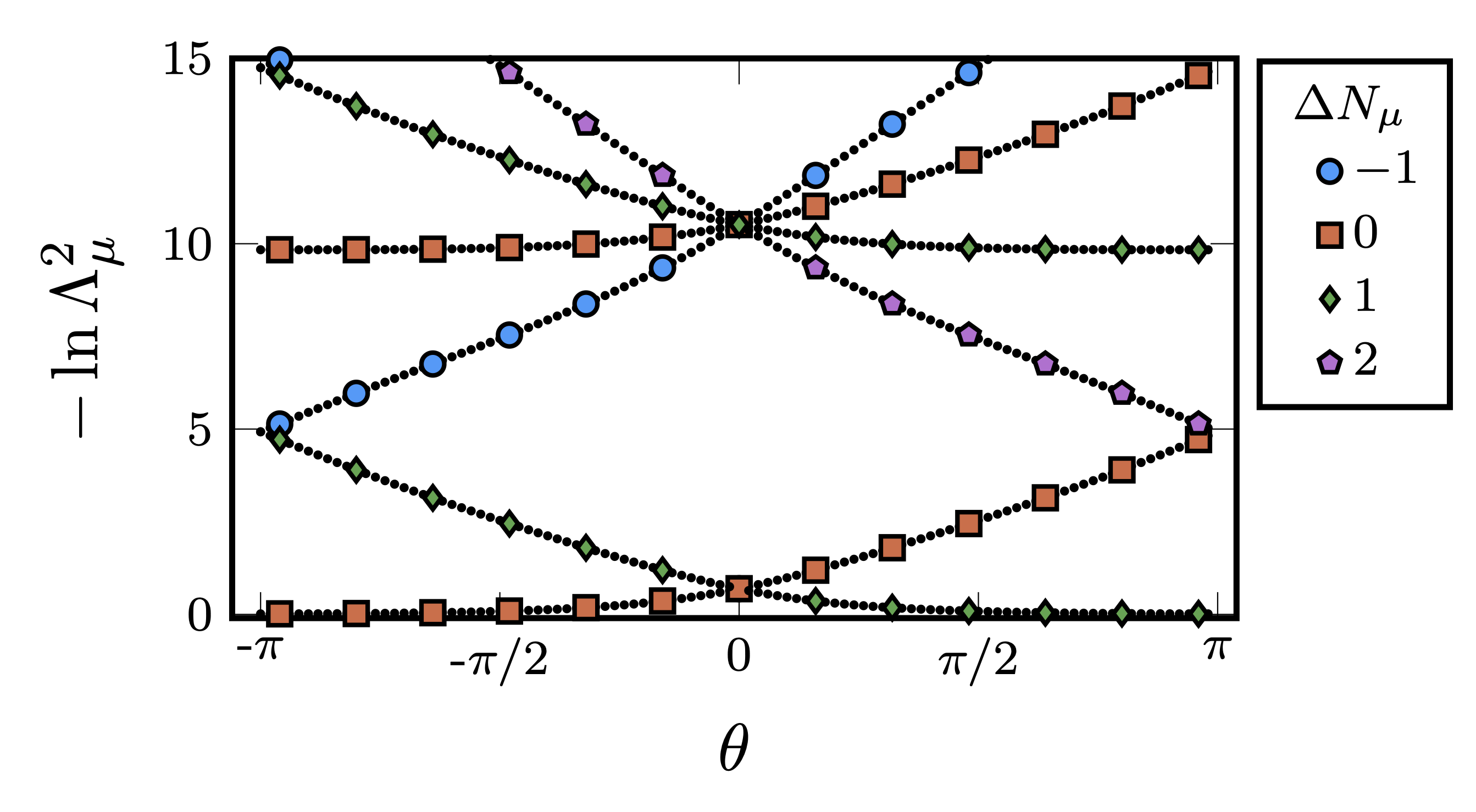}
  \mycaption{Entanglement spectrum of the Rice-Mele model.}
  { We plot the entanglement eigenvalues $-\ln \Lambda_\mu^2$, where $\Lambda_\mu$ are the corresponding Schmidt values, versus the pump parameter $\theta$. Results for the pump cycle with
          $R_\delta=0.5$, $R_\Delta=2.3$, and no disorder, calculated from the
          free-fermion solution for $L=400$ (see also \cite{Hayward2018}). The spectra are in general unbounded, but here we
          only show the lowest few eigenvalues, which, in general, contain the important
          topological information \cite{Li2008}.
     The labels $\Delta N_\mu$ (indicated by markers on every eighth data point) correspond to the particle imbalance for each entanglement eigenvalue. Notice that
   entanglement eigenvalues in the spectrum  wind either up or down or have no winding  if $\Delta N_\mu =  \Delta N_0$, where $\Delta N_0$ is the lowest eigenvalue per cycle. After one cycle, the spectrum has the same values, but with all labels increased by 1 (which occurs at $\theta=0$), indicating the pumping of a single charge.}
  \label{fig:clean-ent}
\end{figure}

The eigenvalues $v_\alpha$ are symmetrically spread around $0.5$ and the low-lying eigenstates of the entanglement Hamiltonian are strongly localized at the chosen cut, see \cite{Peschel2009}. We assign a particle-number imbalance $\Delta N_\mu$ to a state defined by $\vec{n}$ as follows: Focusing on the states of the correlation matrix (when restricted to a subsystem) localized on the left side of the subsystem, eigenvalues $v_\alpha <0.5 (> 0.5)$ correspond to states on the left(right) side of the cut and are assigned a label $N_\alpha= -1 (+1)$.
In order to calculate particle imbalances of the full many-body spectrum of $\hat \rho_A$, states on the left(right) side of the cut are counted as unoccupied(occupied) in Eq.~\eqref{eq:fullrho}, which is then evaluated for all possible occupations $\vec n$. The total particle imbalance of the many-body state $\ket{\mu}$ is 
\begin{equation}
  \Delta N_\mu= \sum_{\alpha=1}^{L} N_\alpha n_\alpha.
\end{equation}
The imbalance $\Delta N_0=0$ is assigned to the unique lowest EEV ($\mu=0$)  for $\theta \in \left(-2\pi, 0\right)$. 

Figure~\ref{fig:clean-ent} shows the entanglement spectrum of the clean system ($w/J=0$) over the course of one
pump cycle ($R_\delta=0.5$, $R_\Delta=2.3$) \cite{Hayward2018}.  
Under an adiabatic modulation of the pumping parameters, the entanglement spectrum shows a continuous flow as long as the system is gapped.
As was discussed in \cite{Hayward2018}, the spectral flow and its nontrivial winding structure is a hallmark of the topological nature of the charge pump.   After one pump cycle, the spectrum returns to itself but the particle imbalances $\Delta N_\mu$ increase by 1, indicating the pumping of one charge across the system.
For instance, the state $\ket{\mu=0}$ with the largest eigenvalue increases its particle imbalance by one (see Fig.~\ref{fig:clean-ent} at $\theta=0$).
The multifold degeneracy of the entanglement spectrum at $\theta=0$ is a result of the chiral symmetry protecting the topological phase of the SSH model.

\subsection{Time-integrated current}
A time dependence is  introduced in Eq.~\eqref{eq:cycle} by parametrizing the pump parameter  $\theta = \theta(t)=2 \pi t /T$ with time.
We drive periodically with a period $T$ by propagating all single-particle states of the lower half of the spectrum via the Crank-Nicholson method (see Sec.~\ref{sec:numerics}).

For the Rice-Mele model, the total current operator is given by:
\begin{equation}
  \hat{J} = i\sum_{j=1}^{L}(J_{j}\hat{c}^\dagger_j\hat{c}_{j+1} -  \hc),
  \label{eq:inttotalcurr}
\end{equation}
identifying the $(L+1)^{\rm th}$ with the $1^{\rm st}$ site, realizing periodic boundary conditions. In praxis, we compute $\langle \hat J \rangle/L$, which should yield $\Delta Q=1$ when 
integrated over one pump cycle. 
This  is a physical quantity that allows us to connect to experiments. 
The expectation value of the ground-state current is identically zero. 
The quantized pumping arises from the virtual admixtures into the higher bands, which are proportional to $O(1/T)$, where $T$ is the pump period. 
In this way, the integrated current over one pump cycle converges to a constant in the $T\gg J$ limit.

The effect of non-adiabatic pumping has been studied in several works, see, e.g.,  \cite{Privitera2018,Kuno2019}. 
For sufficiently slow pumps, it has been shown that the corrections to the exact quantization scale according to $O(1/T^2)$ \cite{Privitera2018}. 

\subsection{Numerical methods}
\label{sec:numerics}
For the static calculations, we directly diagonalize the single-particle Hamiltonian at each $\theta$. For the discretization of the LCM, we take $d_\theta/2\pi=10^{-4}$.
For the real-time simulations, we propagate all single-particle eigenstates of $\hat H(t=0)$ that are populated at half filling. The propagator is approximated for each time $t$ via the Crank-Nicholson method (see, e.g., \cite{Manmana2005} and references therein)
\begin{equation}
  \hat U(t,dt)=\left( 1+ i \hat H(t) dt / 2 \right)^{-1}\left( 1- i \hat H(t) dt / 2 \right)+\mathcal O (dt^3),
\end{equation}
with a time step of $dt J=0.05\:$. The full propagator then reads $\hat U(T)=\prod_{t_s} \hat U(t_s,dt)$ with $t_s=dt\; s,\;s\in{0,1,...,T/dt}$.

\section{Results}
\label{sec:random}

In the main text, all calculations are carried out using the parameters $(R_\delta= 0.5, R_\Delta= 2.3)$. In appendix \ref{sec:appendix}, a subset of the calculations have been carried out using the parameters from \cite{Wauters2019}. A comparison between both parameter sets yields no discernible change in the critical disorder strength.

\begin{figure}[tb]
  \centering \includegraphics{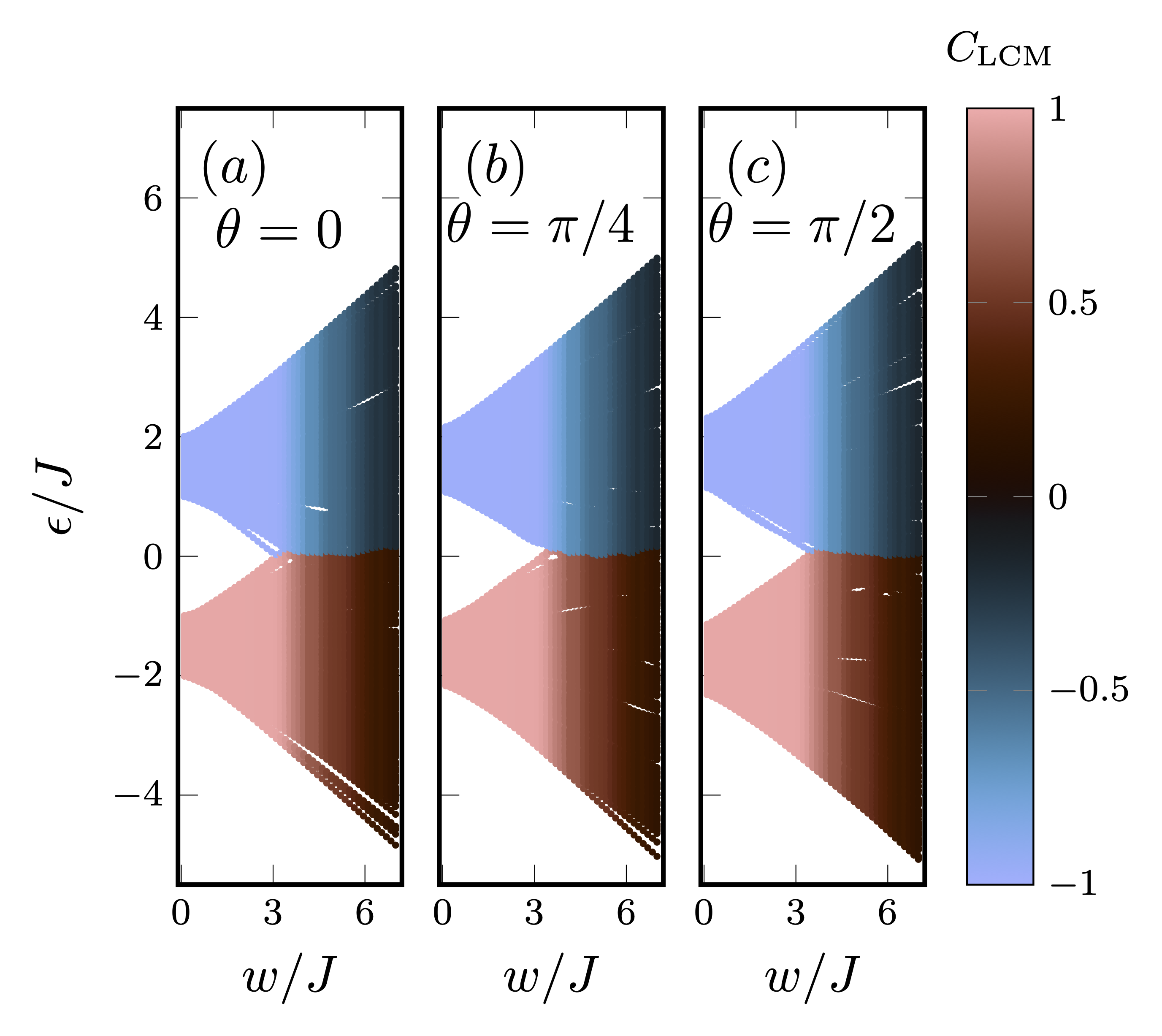}

  \mycaption{Spectrum of the disordered Rice-Mele model for periodic boundary conditions as a function of  increasing disorder strength.}
    {We plot the spectrum for a single disorder realization for $L=400$ and for three values of the pump angle $\theta$.
      The bands are colored based on $\clcm$ at half filling. Edge states do not appear, due to the use of periodic boundary conditions.}
  \label{fig:spec}
\end{figure}

\subsection{Energy Spectra}

We will  compare our measures for the topological properties  to the spectral properties of the static Hamiltonians of the charge pump.
The common notion is that the topological invariant cannot change while the spectral bulk gap persists as $w$ increases.
The single-particle spectrum for  $(R_\delta= 0.5, R_\Delta= 2.3)$ is shown in Fig.~\ref{fig:spec} for various points along the pump cycle for a single disorder realization. 
For this realization, we see a gap closing at $w/J\approx 3.14$ and $\theta=0$. The color indicates the Chern number $\clcm$ of each band calculated via the integrated local Chern marker at half filling (see Secs.~\ref{sec:LCM_method} and \ref{sec:LCM_results}). As long as the bands are clearly separated, $\clcm$ is quantized.

The gap closing can occur at any point in the pump cycle. Relevant information about a topological breakdown is therefore encoded in the minimum gap over the entire pump cycle:
\begin{equation}
 \Delta \epsilon_{\rm min}:= \operatorname{min}_\theta \Delta\epsilon(\theta)/J.
  \label{eq:mingap}
\end{equation}

\begin{figure}[tb]
  \centering \includegraphics{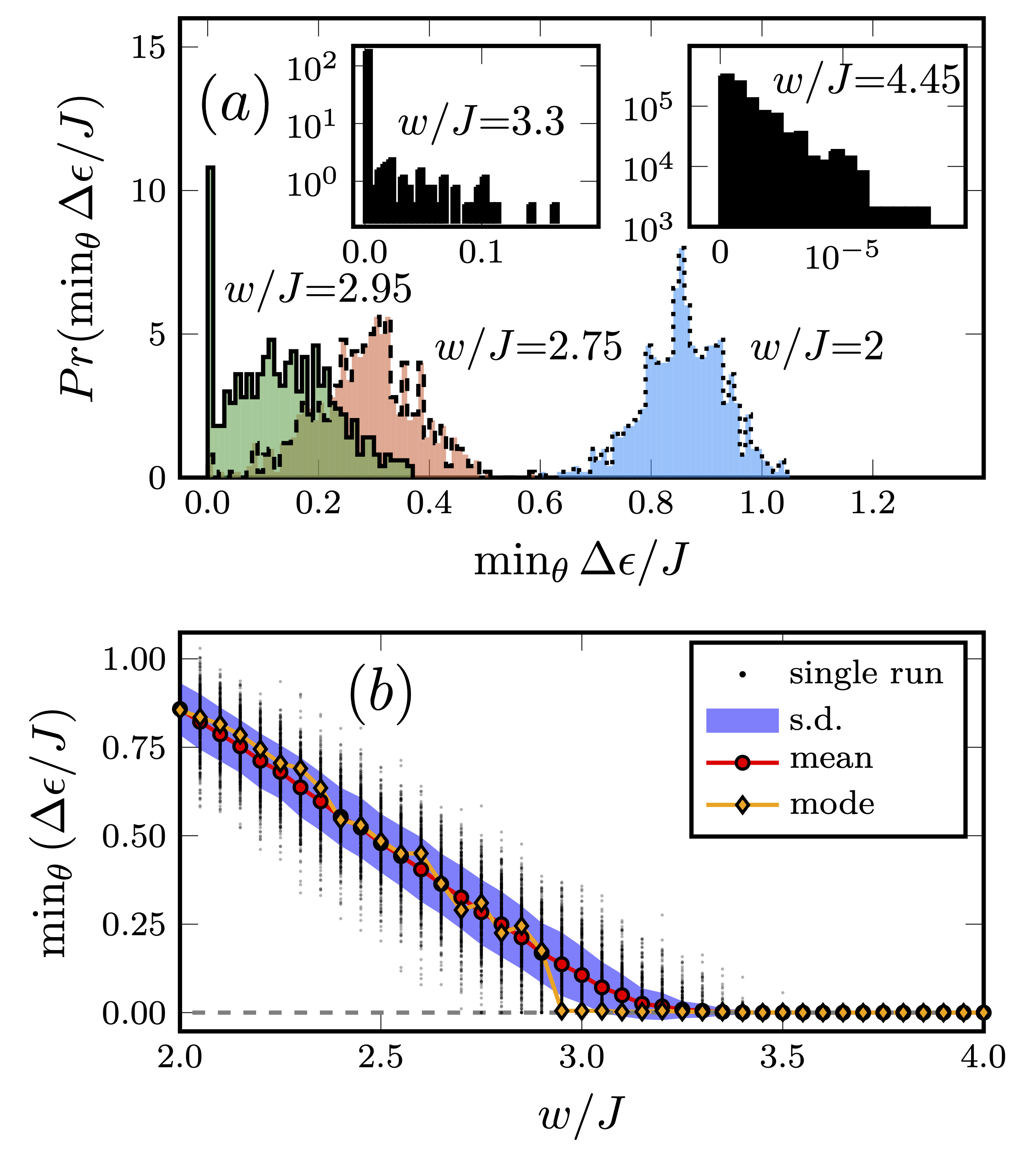}
  \mycaption{Energy gap for diagonal disorder.}
  {(a): Main panel: Distribution of the minimum energy gap along the pump cycle, $\min_\theta \Delta\epsilon/J$, for $w/J=2,\;2.75,\;2.95$ over $n=500$ disorder samples. The most likely value, the  mode, for the minimum gap becomes zero at around $w/J\approx2.95$. Insets: Minimum gap distributions for $w/J=3.3,\;4.45$ and smaller bin sizes. (b): We show the magnitude of  the bulk gap  as a function of  $w/J$. We plot the minimum value found along the whole pump cycle. The red circles and shaded region correspond to the mean and the standard deviation, respectively, computed for  $L=400$. The orange diamonds indicate the mode, which vanishes at $w/J\approx 2.95$. We use a linear $\theta$-grid of $d\theta/2\pi = 0.0001$ that determines the accuracy of finding the minimum gap in a given realization. The dashed line corresponds  to a vanishing gap.
}\label{fig:energy-gap-dist} 
\end{figure}

Figure~\ref{fig:energy-gap-dist} shows the distribution of the minimum gap over $n=500$ disorder realizations for $w/J=2,\;2.75,\;2.95,\;3.3,\;4.45$. For $w/J=2,\;2.75$, the distributions are symmetric and their mean is a meaningful quantity. For $w/J=2.95$, most realizations have a minimum gap of zero, which leads to a transition to exponential distributions from this point onward. These distributions have a maximally likely value, the  
mode, of the minimum gap of zero. The transition to an exponential distribution is exemplified by the insets in Fig.~\ref{fig:energy-gap-dist}(a). Notice that in the insets, different bin sizes are used.
In Fig.~\ref{fig:energy-gap-dist}(b), we show the minimum gap as a function of disorder strength $w/J$. Black dots indicate single realizations. The red circles and blue ribbon denote the mean and standard deviation over all realizations, respectively. The orange diamonds show the mode of the minimum gap distributions. For every $w$, we use a different disorder seed to generate the onsite potentials to preempt a dependence on single realizations. 
While the mean gap closes gradually at around $w/J\approx 3.25$, the mode of the gap shows a very sharp transition to zero at $w/J=2.95$. We argue that the mode and not the mean gap is a meaningful quantity to predict the topological transition of the local Chern maker and the pumped charge, which will be substantiated in Sec.~\ref{sec:compare}.

Since the single-particle gap of the clean system at $\theta=\pi/2$ is $\Delta$, there is, in principle, a disorder realization for which the gap closes at that point for $w \geq \Delta$, where the disorder realization is exactly the (oppositely) staggered potential with strength $-\Delta$.  A single, individual realization, however, has measure zero and hence is expected to become irrelevant in the thermodynamic limit and for the disorder average  over finite systems.

The minimum band gap in the absence of  disorder is $\Delta\epsilon=2J$ at $\theta=0$ for the chosen cycle and parameters used in this paper,  whereas the critical disorder strength needed to close the mode is  $w/J=2.95$. 
This illustrates that the disorder strength can in fact exceed the gap-width of the  single-particle spectrum without spoiling the quantized pumping behavior.

\begin{figure}[t]
  \centering \includegraphics{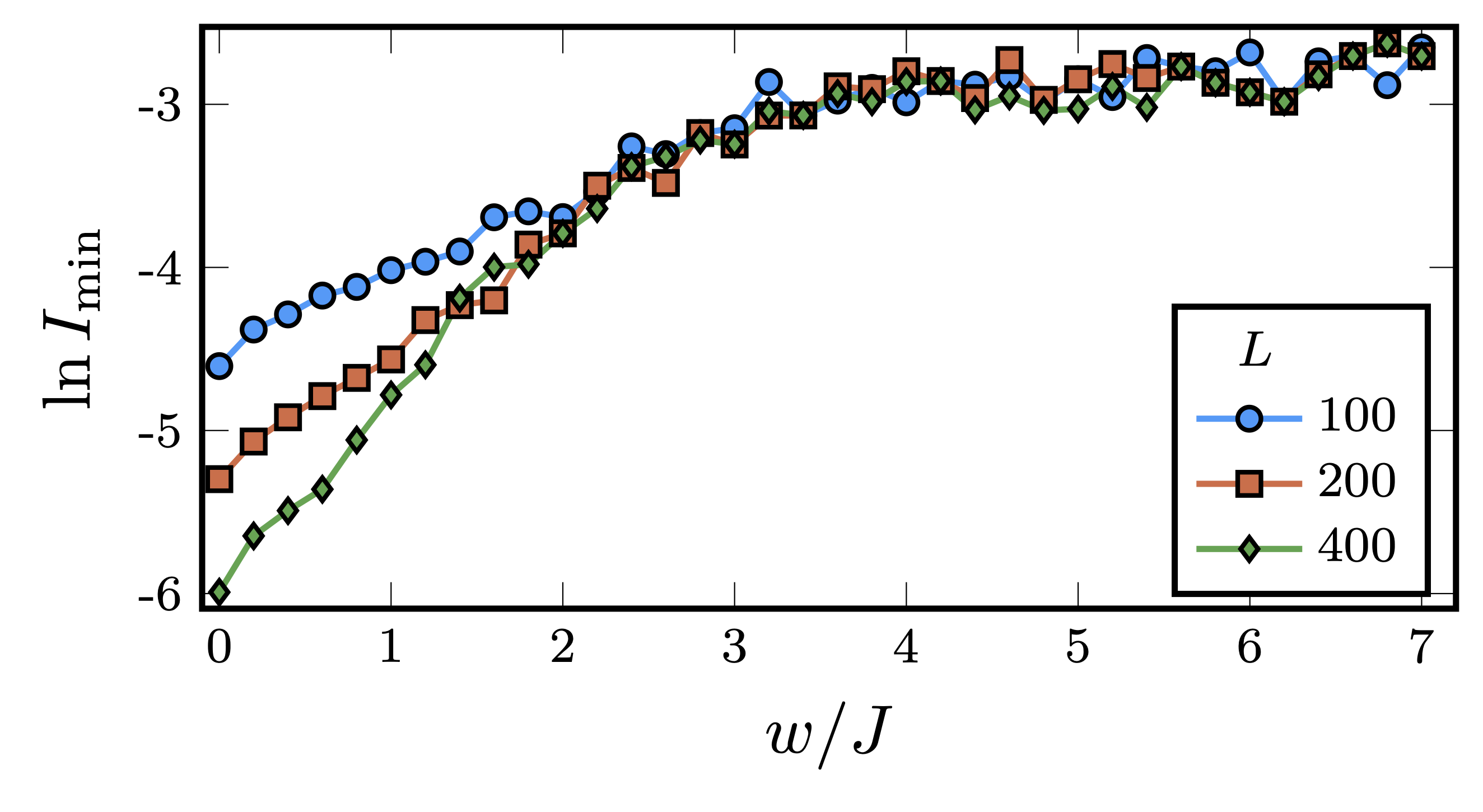}
  \mycaption{Minimum inverse participation ratio $I_{\rm min}$ for diagonal disorder. }{We take the minimum IPR over $\theta$ and over 40 disorder realizations as a function of $w$, for various system sizes.
At $w/J\approx 1.5$, $I_{\rm min}$ becomes independent of system size for systems $L \geq 400$,  indicating that all single-particle states are localized with a localization length much smaller than system size.}
\label{fig:diag-ipr-vw}
\end{figure}

\subsection{Localization of single-particle states }

To demonstrate that our numerically simulated system sizes are sufficient to capture the localized nature of the single-particle states, we compute the inverse participation ratio. 
 Full localization  at any $w>0$ is the expected behavior in one dimension for random diagonal disorder \cite{anderson_absence_1958}.

The inverse participation ratio $I$ is a means of quantifying the degree of localization for a state:
\begin{equation}
  I(\ket{\psi}) = \frac{\sum_{i}  |\braket{\psi|i}|^4 }{\braket{\psi|\psi}}\,.
\end{equation}
Here, $I$ is calculated in the real-space  basis, where $\ket{\psi}$ is a single-particle eigenstate and $\braket{\psi|i}$ is the amplitude on a site $i$. For a completely localized state that has weight on only one site, $I=1$, whereas for a totally delocalized one, $I=1/L$, where $L$ is the system size.

Figure~\ref{fig:diag-ipr-vw} shows the minimum value $I_{\mathrm{min}}:=\min_{\theta,s}I$  of the entire spectrum over one pump cycle and over all given disorder realizations indexed by $s$, corresponding to the most delocalized state. 
The observed behavior is compatible with the expectation of all states being localized for any finite disorder strength. At low disorder, 
 there are states which appear delocalized across the entire system because of the localization length exceeding the system size, indicated by the $L$ dependence of $I_\mathrm{min}$. At higher disorder strengths, the maximally delocalized state is quite short-ranged, with a length scale that is  much shorter than the system length.

For all finite $w$, $I_{\mathrm{min}}$ is much larger than $1/L$ , which is the typical value for a fully delocalized state. For $w/J>1.5$, $I_\mathrm{min}$ is no longer dependent on system size for $L=400$, indicating a localization length much shorter than system size. Crucially, even in this region, quantized pumping is still possible on finite systems and for the disorder average, which will be shown below.
This agrees with the analysis of \cite{Wauters2019}, where the breakdown of quantized pumping has been linked to a delocalization-localization transition of Floquet eigenstates,
which occurs deep in the regime of localized single-particle Hamiltonian eigenstates.

\begin{figure}[t]

  \centering \includegraphics{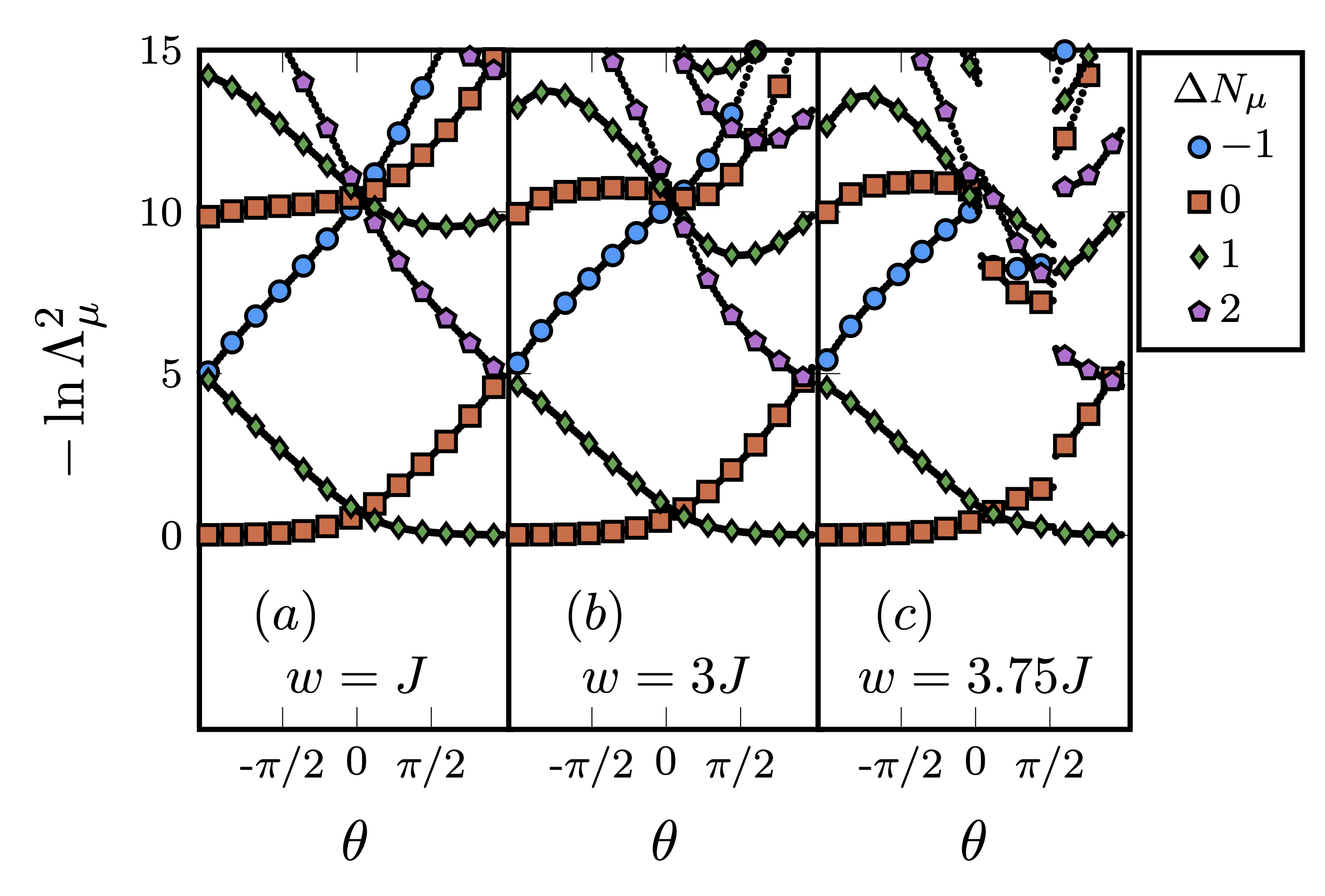}
  \mycaption{Entanglement spectrum for diagonal disorder for several disorder strengths versus pump parameter $\theta$.}{ (a) $w=J$, (b) $w=3J$, (c) $w=3.75J$, computed for $L=400$ sites.
We show data for one fixed disorder configuration and vary $w$. $\Delta N_\mu$ (markers drawn on every eighth data point) are the particle imbalances for each entanglement eigenvalue $-\ln\Lambda_\mu^2$. $\Lambda_\mu$ is the corresponding Schmidt value.}
  \label{fig:diag-ent}
\end{figure}

\subsection{Entanglement spectrum}

Disorder in the onsite potentials breaks the chiral symmetry at $\theta = 0, \pi$ which lead to the topological and the trivial phase of the SSH model, respectively. We see in Figs.~\ref{fig:diag-ent}(a) and \ref{fig:diag-ent}(b) that, for small disorder, the entanglement spectrum is perturbed as $w$ becomes finite, but the topological winding structure of the states is preserved.
The degeneracies in the entanglement spectrum persist at weak disorder yet, since chiral symmetry is broken, they are shifted to
arbitrary positions along the pump cycle, depending on the specific  disorder realization.
At large disorder strength,  there are discontinuities in the entanglement spectrum as exemplified by the data in Fig.~\ref{fig:diag-ent}(c). 
These signal the breakdown of quantized charge pumping.  The position and number of these discontinuities is highly dependent on the particular disorder realization chosen. Therefore, from the entanglement spectrum, it is more cumbersome 
to extract  the exact transition at the breakdown of the quantized pumping than for other measures discussed here.

\begin{figure}[tb]
  \subfloat{\includegraphics{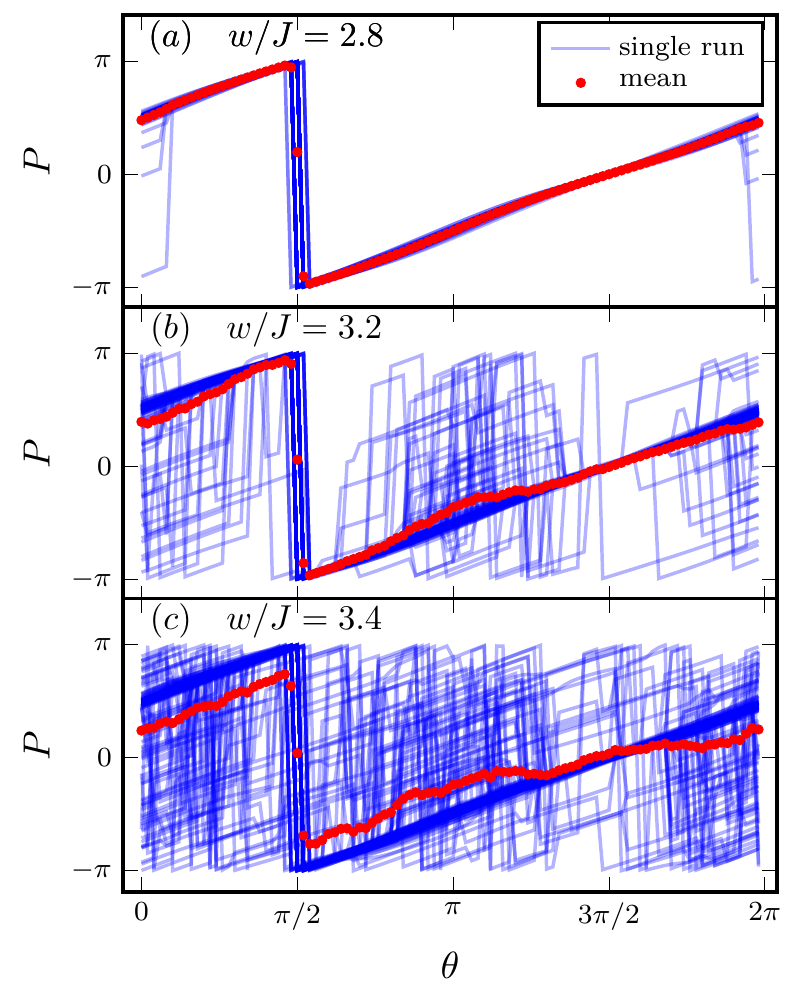}}\\
  \mycaption{Polarization $P$ versus pump parameter for diagonal disorder.}{For each value of the disorder strength  $w$, we show the polarization for 40 disorder realizations as a function of the pump parameter $\theta$ for $L=400$. $2\pi$ discontinuities appear near $\theta=\pi/2$ for $w/J=2.8$ which lead to discontinuities in the mean. For $w/J=3.2$ and $w/J=3.4$, the $2\pi$ discontinuities appear at various points along the pump cycle, which leads to an apparent but misleading continuous mean polarization.}
  \label{fig:diag-pol-conv}
\end{figure}

\begin{figure}[tb]
  \subfloat{\includegraphics{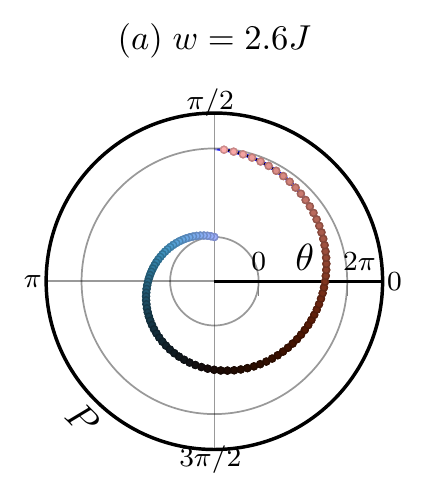}}
  \subfloat{\includegraphics{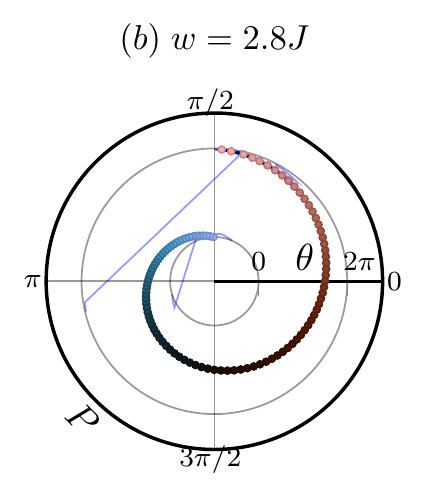}}\\
  \subfloat{\includegraphics{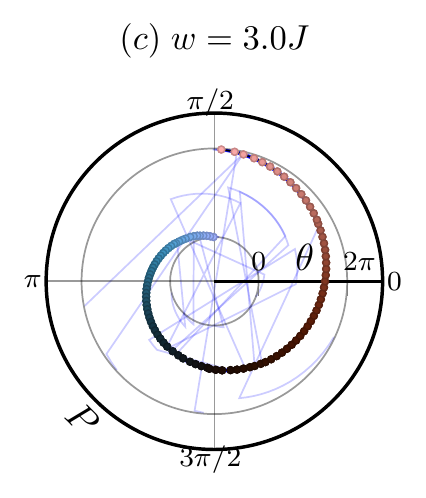}}
  \subfloat{\includegraphics{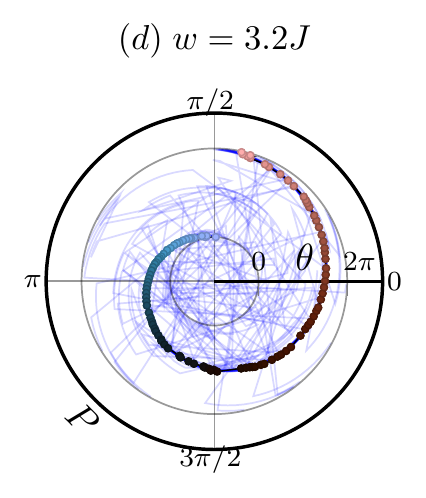}}\\
  \subfloat{\includegraphics{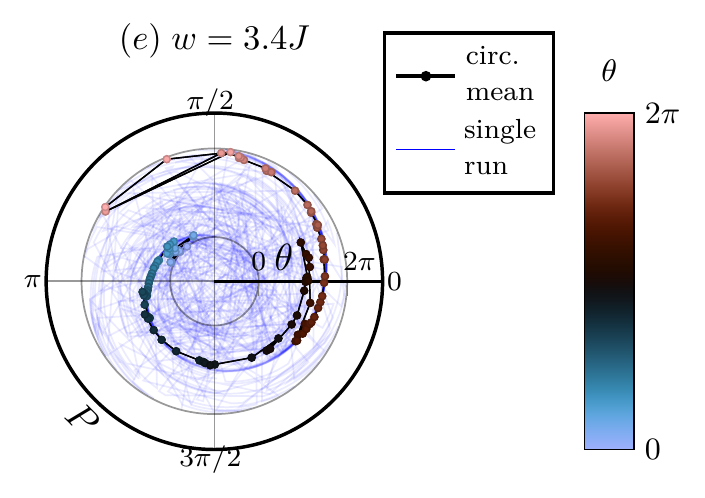}}
  \mycaption{Polarization $P$ versus pump parameter for diagonal disorder.}{ For each value of the disorder strength  $w$, we show the polarization as an angular variable for 40 disorder realizations against the pump parameter $\theta$ for $L=400$.}
  \label{fig:diag-pol}
\end{figure}

\subsection{Polarization}

Figures~\ref{fig:diag-pol-conv} and \ref{fig:diag-pol} show data for the polarization $P$, calculated via Eq.~\eqref{eq:pol-MB2} for $n=40$ disorder realizations as a function of the pump parameter $\theta \in [0,2\pi)$ for various disorder strengths. Due to $P$ being defined modulo $2\pi$, it is convenient to use a polar plot with $P$ being the angular variable and $\theta$ being the radial one. In fact, this is a natural representation, as the $P$ values lie on a circle.

In earlier works, the polarization has been utilized with open boundary-conditions and directly linked with bulk-boundary correspondence \cite{Imura2018}. Here, we employ periodic boundaries. First, we discuss the conventional way of plotting $P$ as a function of $\theta$, which is seen in Fig.~\ref{fig:diag-pol-conv}. Complications arise due to the disorder when performing naive disorder averages: For $w/J=2.8$, the polarization for each sample shows a $2\pi$ discontinuity that stems from the definition of the polarization. These $2\pi$ discontinuities lie near $\theta=\pi/2$, the exact position of which is slightly shifted due to disorder. This leads to a mean polarization that is close to zero (see Fig.~\ref{fig:diag-pol-conv} (a) around $\theta=\pi/2$). Physically relevant discontinuities due to gap closings appear around $\theta=0$ ,or wherever the first gap closing occurs along the pump cycle. Ideally, in the disorder average, the $2\pi$ discontinuities should not lead to a discontinuity in the mean (red dots), which, however, is the case here. For $w/J=3.2,\;3.4$, this issue is amplified, as additional $2\pi$ discontinuities appear for various $\theta$.
Due to these $2\pi$ discontinuities, the naive disorder averaged polarization appears more continuous than it should (see Fig.~\ref{fig:diag-pol-conv} (c) between $\theta=3\pi/2$ and $\theta = 2\pi$).

Plotting $P$ as an angular variable instead, as in Fig.~\ref{fig:diag-pol}, circumvents this problem, as points near $\pi$ and $-\pi$ are close to each other. The colored dots show the circular mean
\begin{equation}
  \overline{P}(\theta)=\operatorname{atan2} \left(\frac{1}{n} \sum_{j=1}^{n} \sin P(\theta)_{j}, \frac{1}{n} \sum_{j=1}^{n} \cos P(\theta)_{j}\right),
\end{equation}
which is the natural averaging procedure for points on a circle, for each pumping time $\theta$, where atan2 is the two-argument arctangent. Single realizations are shown as blue lines.
For low disorder strengths, all realizations lie on top of each other and the circular mean exhibits a smooth dependence on $\theta$ with winding 1. From $w/J=2.8$ onward, some realizations display discontinuities as indicated by the jumps in the blue lines. At $w/J=2.8$, only a few realizations have such jumps, whereas for $w/J=3.4$, almost all polarizations are discontinuous, with individual realizations showing multiple discontinuities of arbitrary size.
The circular mean becomes visually discontinuous around $w/J \approx 3$, indicating that the quantization breaks down in a large number of realizations. Comparing the regular mean to the circular mean in Figs.~\ref{fig:diag-pol-conv}(c) and \ref{fig:diag-pol-conv}(e) shows that the former exhibits no such discontinuities except the trivial one at $\theta=\pi/2$.
We associate these discontinuities in the circular mean with the breakdown of
 quantized charge pumping.
In this case, the winding number of the polarization is ill-defined: It is impossible to know whether the jump in the polarization is clockwise or counterclockwise for single realizations. 
Despite this fact, the circular mean retains a sense of direction in its winding and winds around exactly once, meaning that on average, charge is transported across the system in a well-defined direction. For a topologically trivial pump cycle, the circular mean shows no winding, as expected (data not shown). This suggests that the circular mean of the polarization 
still preserves information on the quantized pumping  inherited from the neighboring topological phase even beyond the critical disorder strength.

\begin{figure}[tb]
  \centering \includegraphics{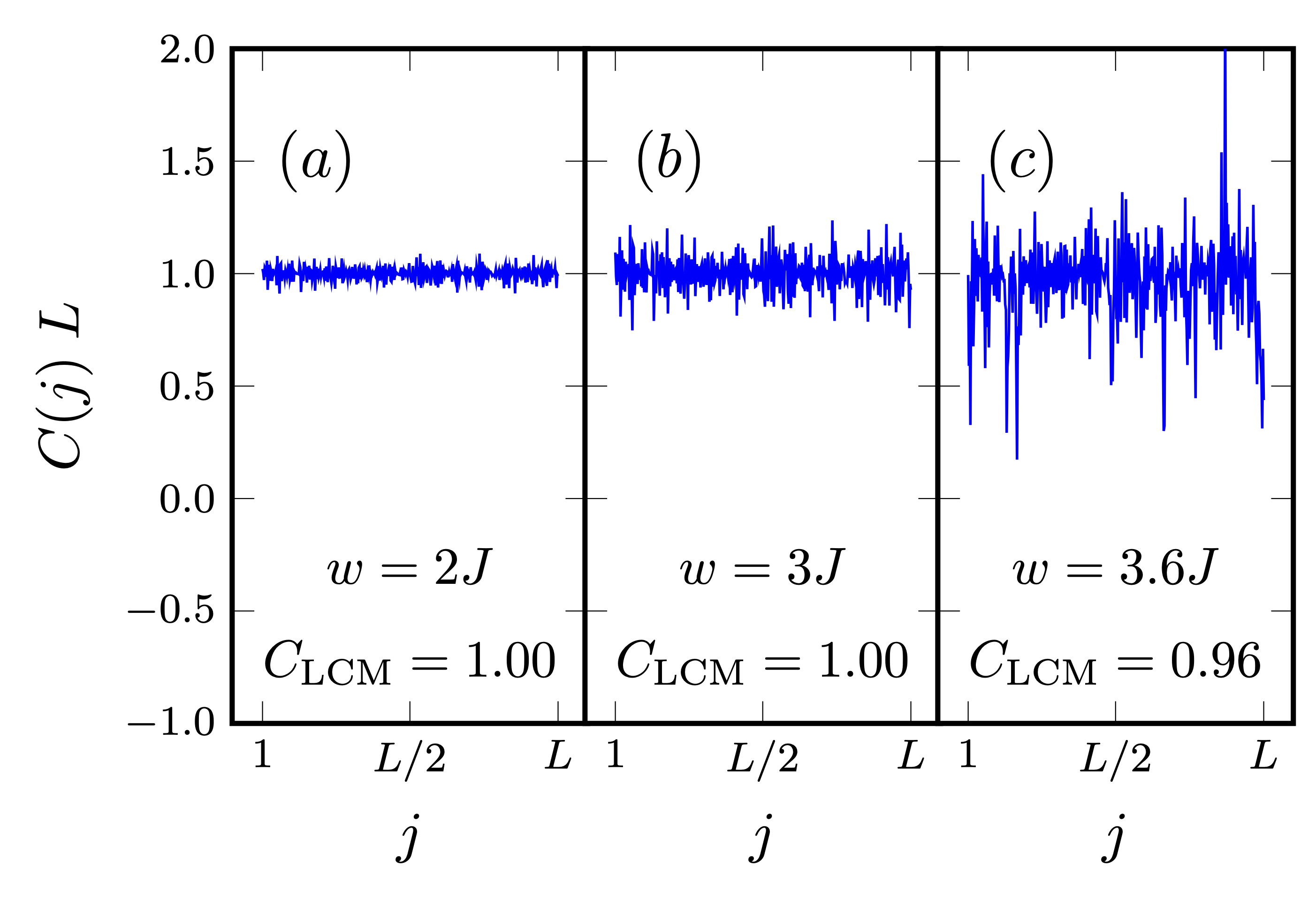} 
  \centering \includegraphics{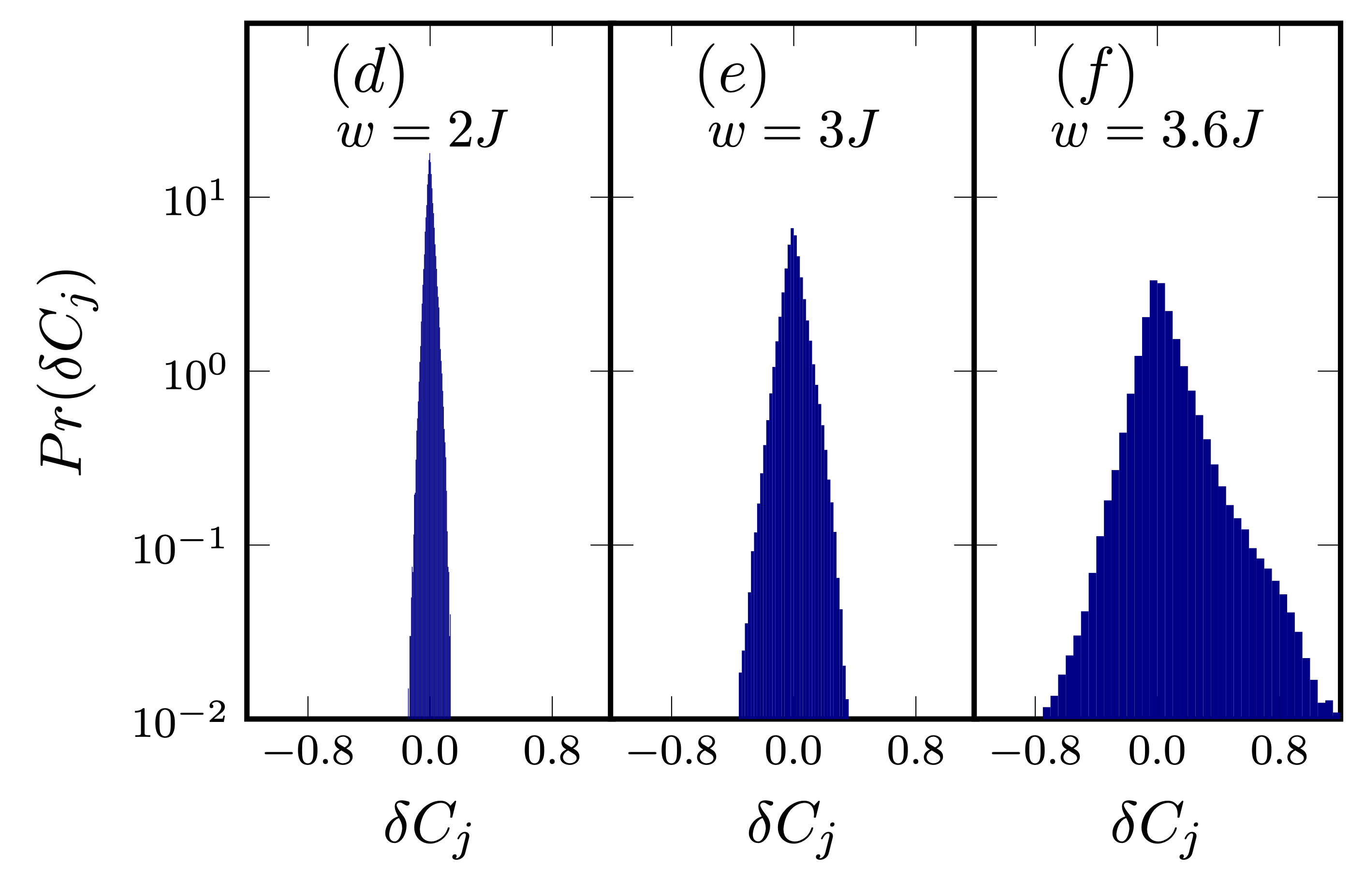}
  \mycaption{Local Chern marker.}{(a)-(c): $C(j)L$ versus position $j$ for a single realization of diagonal disorder and for disorder strengths (a) $w=2J,
\clcm=1.00$,(b) $w=3J, \clcm=1.00$, and (c) $w=3.6J, \clcm=0.96$, computed for $L=400$. 
(d)-(f): Normalized distribution of local Chern-marker fluctuations $\delta C(j)=  1 - C(j)\;L$ for (d) $w=2J$, (e) $w=3J$, and (f)  $w=3.6J$ from 500 disorder realizations for $L=400$. Note that the distribution in (f) is slightly skewed towards positive values, see Fig.~\ref{fig:skew-lcm}.
}
\label{fig:diag-lcm}
\end{figure}

\begin{figure}[tb] 
  \centering \includegraphics{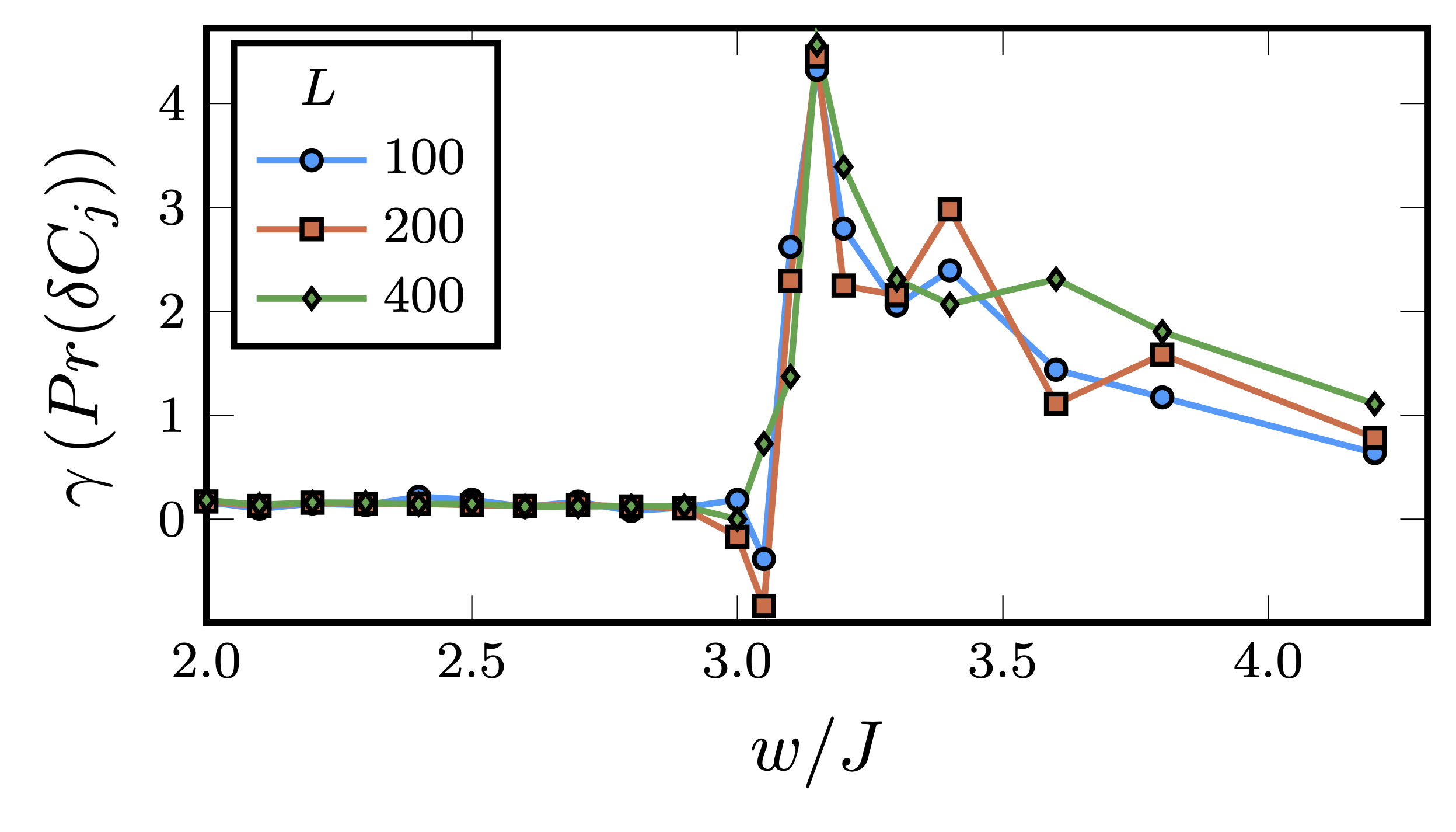}
    \mycaption{Transition from distributions of local Chern marker fluctuations.}{
Skewness $\gamma$ of the distribution of local Chern number fluctuations $Pr(\delta C(j))$ versus disorder strength for $L=100,200,400$. Computed from 500 disorder realizations for $w/J\geq3$ and from 100 realizations for $w/J<3$.
}
\label{fig:skew-lcm}
\end{figure}

\subsection{Chern number from integrated local Chern marker}
\label{sec:LCM_results}

{\em Local Chern marker and its fluctuations.} The local Chern marker $C(j)$ is shown for $w/J=2,3,3.6$ in Figs.~\ref{fig:diag-lcm}(a)-(c) for a single disorder realization. For zero disorder (data not shown here), the local Chern marker is $C(j)=1/L$ (within our numerical accuracy) as expected. As $w$ increases, fluctuations
around this value emerge and increase as $w$ grows. 
 For $w/J=3.6$, integer-quantization clearly no longer holds. 

We next demonstrate that the spatial fluctuations of the local Chern marker defined via
\begin{equation}
\delta C(j) = 1 - C(j)\;L \label{eq:cj_fluct}
\end{equation}
contain relevant information and can be used to pinpoint the transition. 
A local Chern marker $L C(j)>1$ $(\delta L C(j) <0)$ implies that an excess amount of charge crosses site $j$ per pump cycle, while 
for $L C(j)<1$ $(\delta L C(j) >0)$ a lesser amount moves through. In order for the system to possess an integer-quantized Chern number,
these fluctuations have to cancel out after summing over the whole sample. Our analysis shows that this happens for individual realizations in the topological regime. 
 
The respective distributions of $\delta C(j)$ obtained from $n=500$ disorder configurations are plotted in Figs.~\ref{fig:diag-lcm}(d)-(f).
In the topological regime [see Fig.~\ref{fig:diag-lcm}(d)], $\mbox{Pr}(\delta C(j))$ is symmetric and sharply peaked around zero, while the variance  gradually increases as the transition point is approached.
For $w=3.6J$, the distribution becomes skewed towards $\delta C(j) > 0$.
In order to determine the breakdown of the quantized pumping, we compute the skewness $\gamma$ of the distributions from
\begin{equation}
\gamma =  \frac{\frac{1}{n L} \sum_{i=1}^{Ln } (x_i -\bar x)^3}{  (\frac{1}{L n} \sum_{i=1}^{nL} (x_i -\bar x)^2)^{3/2}  }    \label{eq:skew} \,
\end{equation}
with $x_i$ being $C(j)$ computed for a given disorder realization.
The dependence of $\gamma$ on the disorder strength is shown in Fig.~\ref{fig:skew-lcm} for $L=100,200,400$.
Remarkably, the skewness exhibits a strong signature at the transition: on small systems, it first decreases but then sharply increases. The best estimate from our available data for the transition point is $w/J=3.0$. There is no detectable $L$-dependence for the transition point for the system sizes considered here within the used grid of $\Delta w=0.1J$ around the transition region.
The nonzero skewness can be understood as follows.
As the gap closes, states from the original bands will mix
with those of the lower bands, mixing in a tendency of 
an opposite charge transport. Since the states are localized
due to disorder, this occurs locally, and results in a deficit of 
$C(j)$ and hence locally larger values of $\delta C(j)$, without changing the mode of the $\delta C(j)$ distributions.
We conclude that the analysis of the full distribution of local Chern marker fluctuations and of its moments is very useful to obtain a quantitative picture of the breakdown, as will
be substantiated by the following comparison with the pumped charge from the time-integrated current.

\begin{figure}[t]
\centering \includegraphics{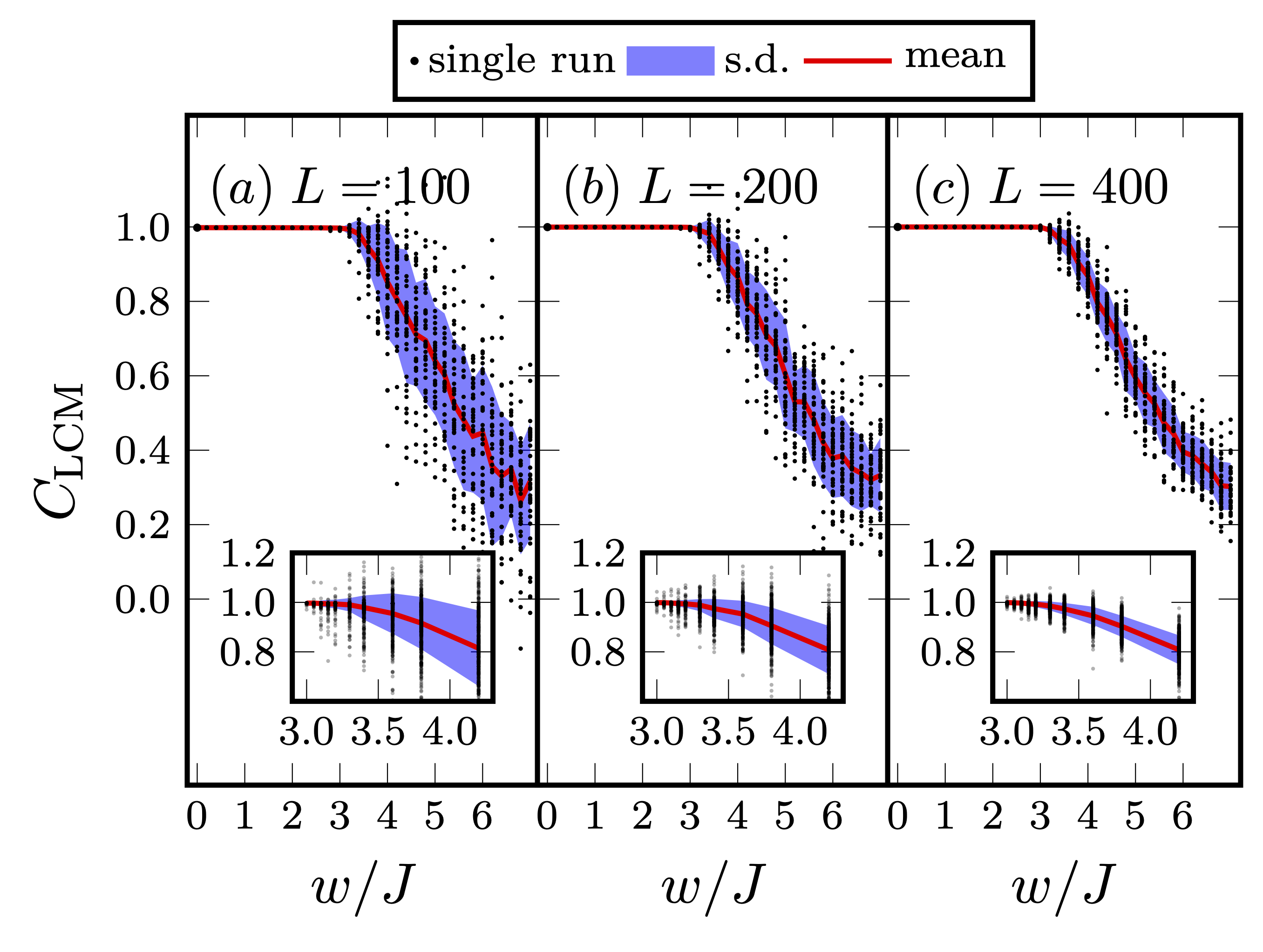}
\mycaption{Sum $\clcm$ of the local Chern marker $\clcm$ for diagonal disorder as a function of $w$.}{(a) $L=100$,  (b) $L=200$, (c) $L=400$. Main panel: 40 samples at each disorder strength. Insets: Zoom into the transition region with 500 samples. The red line indicates the mean value for  each disorder strength and the standard deviation is represented by the shaded region.
}
\label{fig:diag-lcm-dist}
\end{figure}

{\em Integrated local Chern marker and breakdown of quantized pumping.} Figure \ref{fig:diag-lcm-dist} shows the breakdown of the quantized integrated local Chern marker as a function of $w$ for different system sizes $L$. 
Black dots indicate single realizations. The red line and the blue shaded region are the disorder averages and standard deviation, respectively. 

The sum
\begin{equation}
  \clcm=\sum_j C(j)
\end{equation}
and thus the total Chern number remains quantized up to $w\approx 2.3J$ within our numerical accuracy, while for
larger values $2.3J< w\lesssim 3J$, $\clcm$ starts to slowly decrease, due to rare realizations without exact quantization.
An interesting question concerns the system-size scaling of the mean Chern number in the transition region (see the discussion and further references in \cite{Altland2015}).
Since our work is concerned with the properties of ensembles of finite-size systems as realized in quantum-gas experiments, we do not further pursue this direction
and leave this for future research. We further note that finite-size corrections for the pumped charge were studied in \cite{Li2017}.

At strong disorder $w/J\gtrsim  3$, the system becomes gapless and hence the Chern number is illdefined. We therefore expect a breakdown of the quantized charge pumping supported by the data.
Nevertheless, $\clcm$ still spreads around a central mean value over all realizations.
This mean value, however, starts to deviate from $\clcm=1$ and indicates the breakdown of the quantization of the charge pump.
The fluctuations around this mean value
decrease as $L$ increases. 
Note that the standard deviation around the disorder-averaged Chern number $\bar C_{\rm LCM}$ increases just at the point where $\clcm$ ceases to be quantized.

\begin{figure}[tb]
  \centering \includegraphics{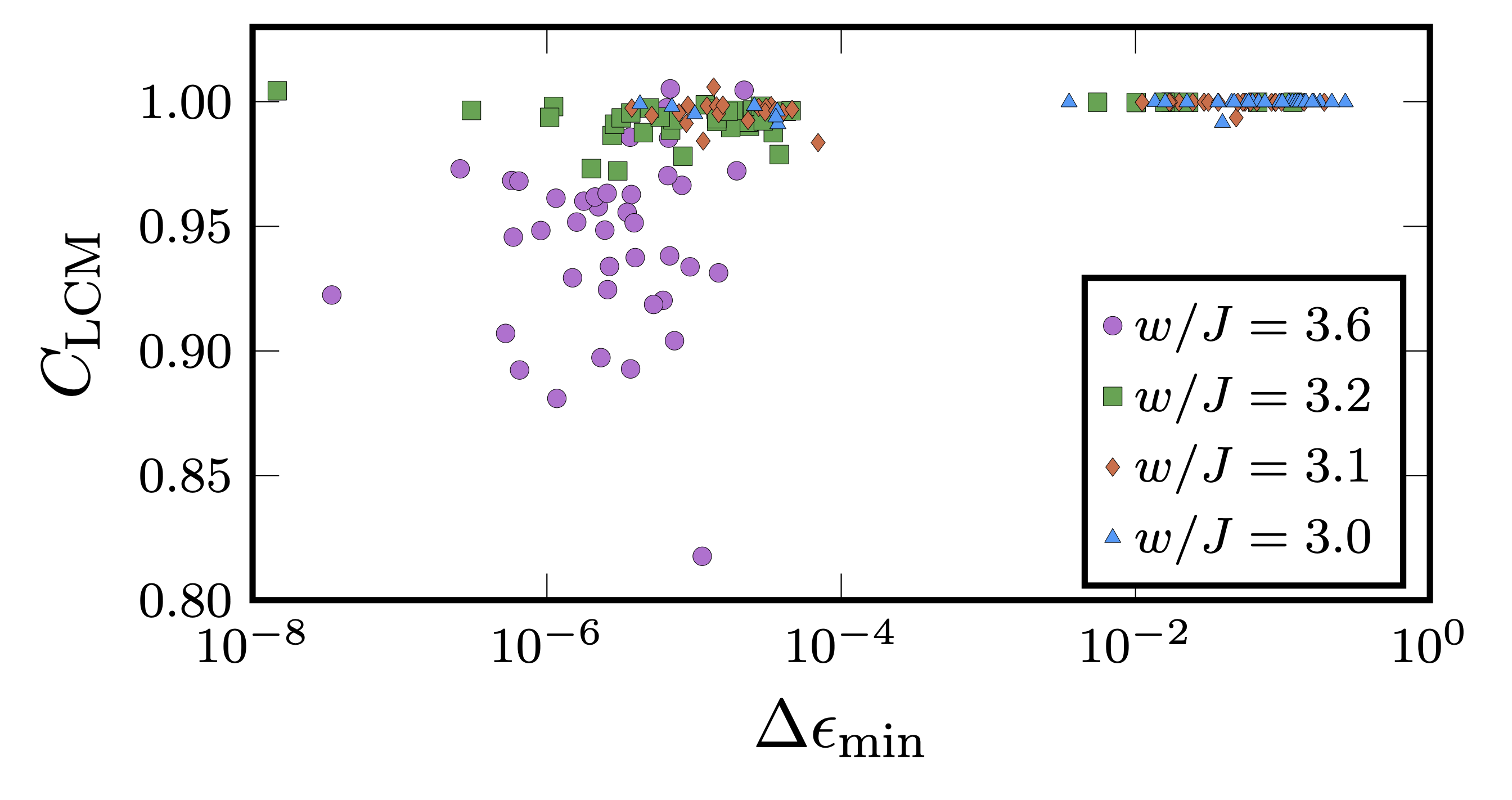}
  \mycaption{Correlation between $\clcm$ and energy gap.} { We compare $\clcm$ to the  minimum global energy gap for 40 disorder realizations for each value of $w$. We see that realizations with non-quantized $\clcm$ lie in the vicinity of vanishing energy gaps that are of the order of $10^{-6}$ in the chosen $\theta$ grid for the energy calculation. These points shift to the left when refining the $\theta$ grid, whereas points on the right remain unchanged (not shown). The spread of $\clcm$ increases for increasing $w$. The two outliers on the right side are atypical realizations with poor convergence for the LCM. All data computed for $L=400$.
 } \label{fig:diag-transitions}

\end{figure}

For single realizations, the notion that a topological quantity cannot change without a gap-closing is confirmed, which is shown in Fig.~\ref{fig:diag-transitions}. We plot the integrated local Chern marker $\clcm$ versus the minimum  gap  min$_{\theta}\Delta \epsilon$ for each realization
and different disorder strengths. Points on the left with min$_{\theta}\Delta \epsilon/J  \lesssim 10^{-5}$ indicate realizations that close the energy gap, with a minimum gap size that scales with the $\theta$-grid size used for calculating the energy. The $\clcm$ for these realizations exhibits non-integer values with a spread that increases for larger $w/J$. 
The points on the right side of the figure correspond to  realizations that do not close the gap and have an integer-quantized $\clcm$. Their minimum gap is invariant under the $\theta$-grid size used for the energy search.  For large $w/J$, the ratio of gap-closing realizations increases sharply. 
We thus conclude that the  breakdown of topological charge pumping on {\em finite} systems
occurs  due to sufficiently many individual realizations acquiring close-to-zero gaps well before the mean (disorder-averaged) minimum gap closes. This
scenario applies to the finite-size quantum-gas experiments where typically, $L \sim 100$, and averages over many one-dimensional systems
are measured.

\subsection{Time-dependent case and integrated current}
\label{sec:random_time}
In this section, we study the dependence of the time-
integrated current Eq.~\eqref{eq:inttotalcurr} on the finite period of the pump cycle. 

In the adiabatic limit, any finite system will ultimately have an exactly quantized integrated current, as the system is gapped. However, the gaps in some systems may become exponentially small in the system size. On the other hand, a finite period $T$ can lead to non-quantized behavior due to Landau-Zener tunneling \cite{Privitera2018}, even in the presence of a band gap. 

We first discuss the evolution of the time-dependent current  as a function of time, for various periods $T$ (data not shown here).
For $T/J=10$, the pumped charge per period is never quantized, due to non-adiabatic excitations to the second band. The quantization of the pumped charge sets in at $T/J=100$ for large values of $w$ in the vicinity of $w=3J$. For such a slow driving, pumping below the critical disorder strength $w/J\approx3$ leads to quantized particle pumping. Beyond that disorder strength, the pumped charge ceases to be quantized.

\begin{figure}[tb]
  \centering \includegraphics{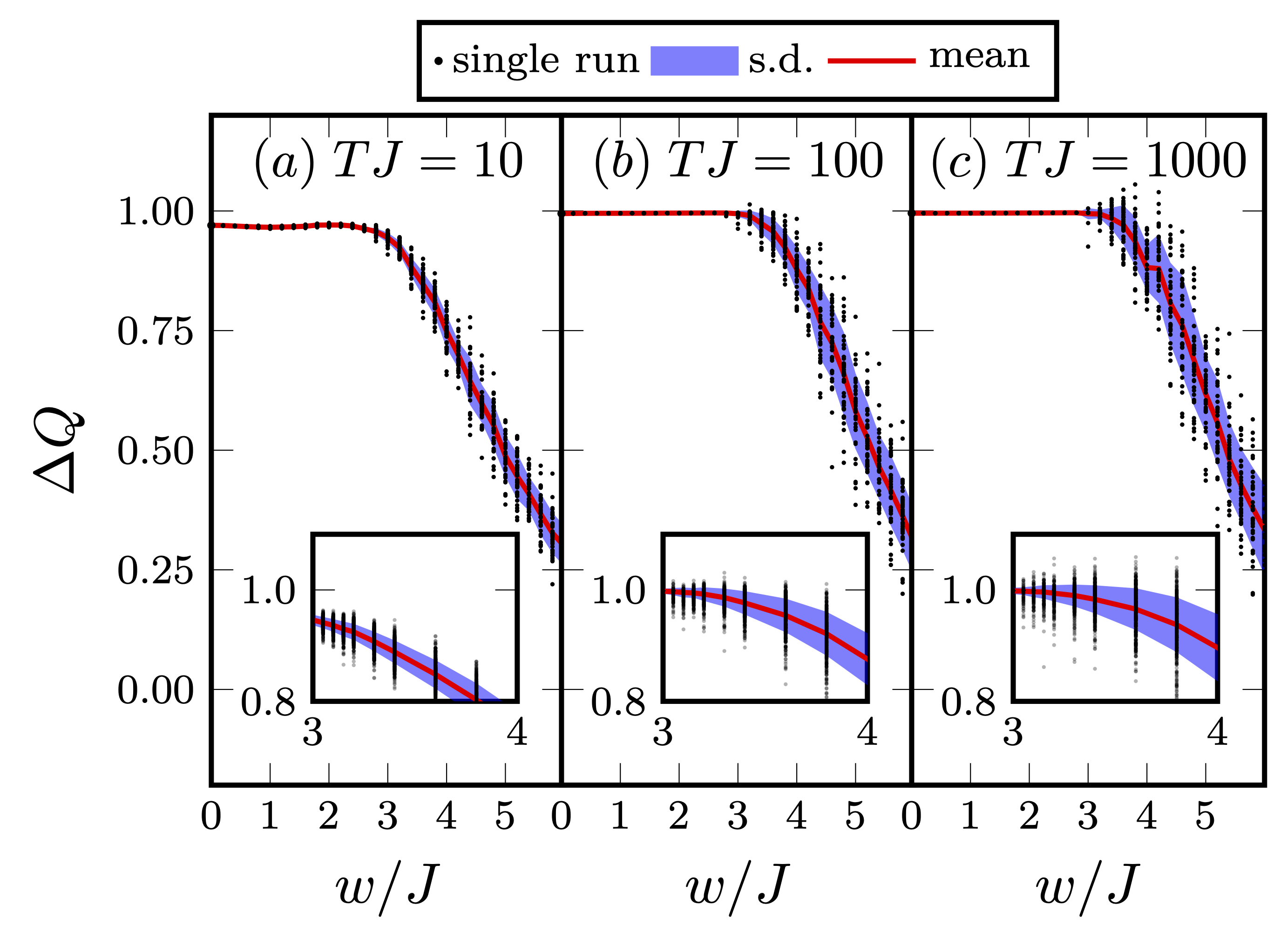}%

  \mycaption{Distribution of  pumped charge computed from the time-integrated current as
  a function of disorder strength.} { Computed for a number of period lengths $T$. Main plots: $n=40$ samples for each disorder point and time-step $dt J=0.1$. Insets: Transition region computed for $n=500$ realizations and $dt J=0.05$ ($L=400$).}
  \label{fig:diag-distribution-time}
\end{figure}

In Fig.~\ref{fig:diag-distribution-time}, we show the distribution of the pumped charge, calculated from the time-integrated current over 500 disorder realizations and for various pumping periods $T$. For $T=10J$, the pumped charge is not quantized, presumably  due to Landau-Zener tunneling into the second band. For $T=100J$ and $T=1000J$, the average pumped charge is similar to what has been found for the Chern number computed via the LCM. Quantization breaks down at the critical disorder strength $w/J\approx3$.

\subsection{Comparison between instantaneous and time-dependent measures}

\label{sec:compare}

\begin{figure}[t]
  \centering \includegraphics{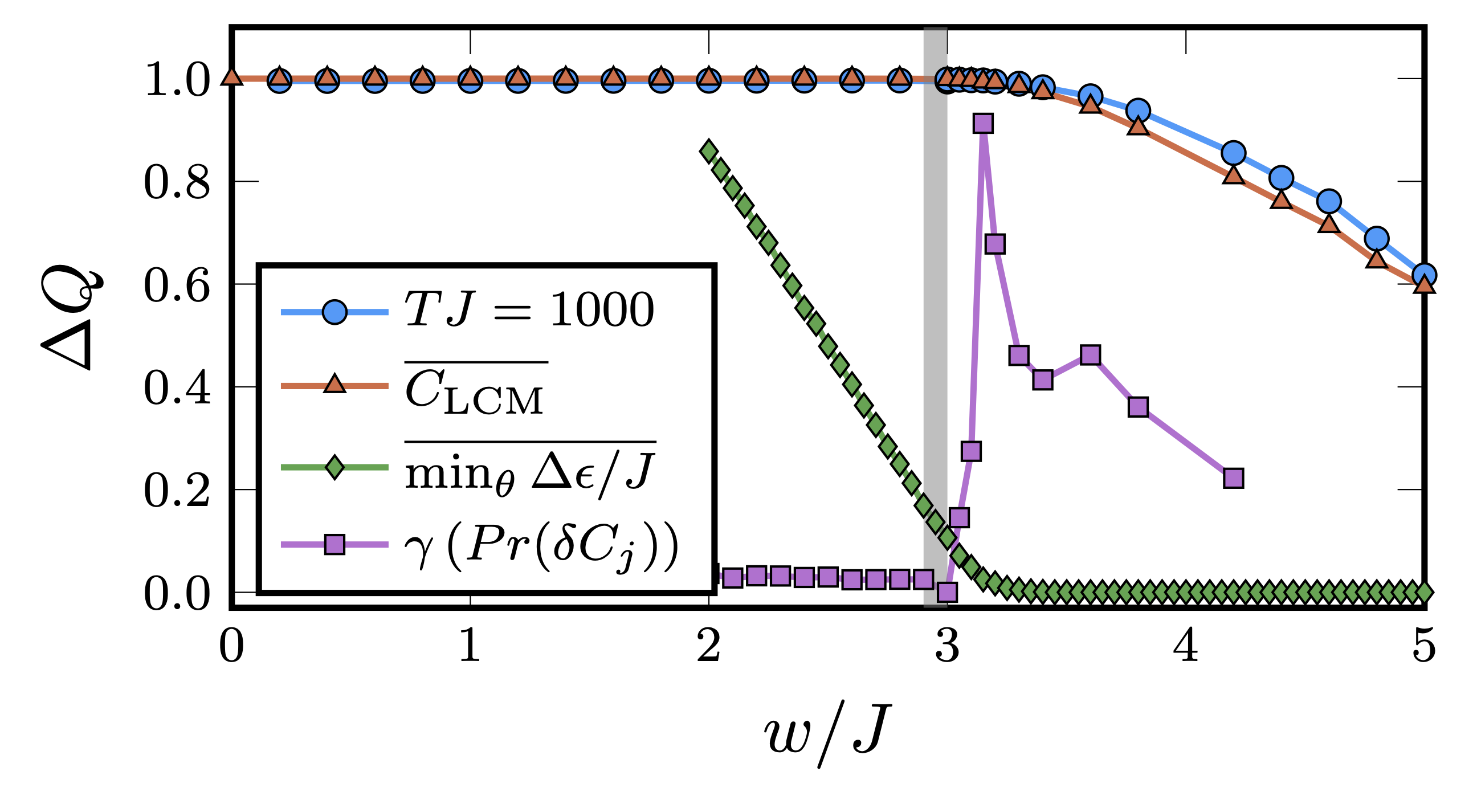}

  \mycaption{Comparison of different measures for the pumped charge.} {We compare $\clcm$,  the average pumped charge computed from the time-integrated current as
  a function of disorder strength, the mean minimum gap found along the pump cycle and the skewness of LCM distributions. The skewness has been scaled down for better visibility. The shaded region indicates the point where the mode of the minimum gap along the pump cycle closes at $w/J\approx2.95\pm0.05$. Similar results were obtained in \cite{Wauters2019} from Floquet states for different parameters. At larger values of $w\gg 8J$, one finds $C=0$ \cite{Wauters2019}. Computed for a pump period of $T=1000J$. $n=500$ ($n=40$) samples were included in the average for each value of $w/J$ for $3<w/J<4.2$ ($w/J\leq 3$, $w/J\geq 4.2$), $L=400$.
 } \label{fig:diag-average-time}
\end{figure}

Figure \ref{fig:diag-average-time} shows a comparison between the disorder average of the minimum energy gap, the deviation of $\clcm$ from $C=1$, the skewness of local Chern marker 
distributions and the time-integrated local current as a function of $w/J$.
We have also studied the parameters from \cite{Wauters2019} using the integrated local Chern marker and consistently observe a breakdown at $w/J \approx 3$, in agreement with \cite{Wauters2019}.

Interestingly, the pumped charge $\Delta Q$ obtained from the time-dependent simulations and $\clcm$ are very similar to each other even
in the trivial regime. This should be taken with some caution, due to the aforementioned numerical difficulty of obtaining converged results from the local Chern marker in the trivial regime for individual realizations (see Sec.~\ref{sec:LCM_method}).

We stress that  the breakdown of the Chern number quantization occurs at $w/J\approx 3$, whereas the mean minimum gap closes at $w/J\approx 3.3$,
while the observed breakdown of integer quantization of $\clcm$ and the pumped charge $\Delta Q$ agrees well with the closing of the typical value of the gap, the mode (vertical shaded region).
This value also agrees well with the onset of a significant skewneess $\gamma$ in the distribution of local Chern markers.
These observations further support our assertion that  the most likely gap, rather than the disorder-averaged gap, should be considered when quantifying the transition point. 

In conclusion, all three quantities computed in the instantaneous basis and the integrated current suggest the stability of quantized charge pumping
at weak disorder. A breakdown is observed for $w/J\gtrsim 3$ and can  best be read off from the skewness of the distributions of local
Chern-marker fluctuations, consistent with the closing of  the most likely energy gap. 
The pumped charge is directly accessible in experiments \cite{Lohse2016,Nakajima2016,Nakajima2020} and should therefore agree with the theoretical predictions for adiabatically slow pumping on finite systems.

\section{Summary and discussion}
In this work, we studied the Rice-Mele model of spinless fermions in the presence of random diagonal disorder.
We used exact diagonalization to compute a set of static measures to characterize the properties of a charge pump, including the 
polarization, the entanglement spectrum, and the integrated local Chern marker. These quantities were  computed in the instantaneous eigenbasis.
We demonstrated that all these measures indicate a breakdown of the quantized charge pumping
at sufficiently strong disorder. 
The breakdown of quantized pumping manifests itself as a breakdown of winding in the spectral flow of the entanglement spectrum. Plotting the 
polarization as an angular variable makes the breakdown particularly transparent in this quantity.
The integrated local Chern marker appears to be the best suited for determining the transition point quantitatively.

In particular,
the fluctuations of the local Chern marker around the bulk Chern number provide relevant information about the breakdown.
It would be very desirable to develop a qualitative interpretation of these fluctuations. For instance, it remains open whether nonlocal adiabatic
processes play a role in topological charge pumping. These effects were described by Khemani {\it et al.} \cite{Khemani2015} as a consequence of adiabatic
variations of a single onsite potential in an Anderson insulator. The situation in a charge pump is not entirely different although there, the variation of
onsite potentials happens in a correlated fashion.

The critical disorder strength obtained from the integrated local Chern marker agrees well with the point at which the bulk gap closes, while for finite systems, we emphasized the importance of sample-to-sample fluctuations.
In particular, we demonstrated that the typical value of the minimum bulk gap along the gap cycle is much better suited to describe the distribution 
obtained from finite systems, rather than the disorder-averaged gap. This may not be surprising, given the known existence of Lifshitz tails at the edges of 
spectra of disordered systems (see, e.g., \cite{Mezincescu1993}).
The topological transition occurs in an Anderson insulator, showing that topological charge pumping is robust against localization, consistent with the results
of \cite{Nakajima2020}.

We complemented the analysis with time-dependent simulations of the time-periodic pump process and observe deviations from integer-quantized
pumping due to a breakdown of adiabaticity for fast pumping. For sufficiently slow pumping, we find agreement with the critical disorder strength obtained from 
the bulk Chern number computed in the instantaneous basis.

In this work, we presented a comprehensive comparison of several measures from the instantaneous basis and direct time-dependent simulations.
Matching the physical picture for local transport processes to the Floquet-localization picture of Ref.~\cite{Wauters2019} would be an interesting
next direction. In this regard, the limit of frequency going to zero in the Floquet picture  might be subtle on finite systems, as one should recover the behavior of the instantaneous-basis behavior
(see \cite{Wauters2019}).

Several studies have already theoretically addressed the question of charge pumping in an interacting system \cite{Berg2011,Rossini2013,Ke2017,Kuno2017,Nakagawa2018,Hayward2018,Qin2018,Mei2019,Stenzel2019,Lin2020,Greschner2020,Lin2020a}, which should be
realizable in state-of-the-art quantum-gas experiments \cite{Lohse2016, Nakajima2016,Schweizer2016,Nakajima2020}.
Conceptually and from a methodological point of view, the question arises which approaches are best suited to compute the Chern number for
a charge pump in a many-body system. The direct calculation of the current in time-dependent simulations  will always work, yet requires making the pump period large.
The Floquet approach of Ref.~\cite{Wauters2019} cannot easily be extended to the many-body case, where most Floquet approaches are based on the high-frequency limit \cite{Eckardt2017}, opposite to
the regime of low frequencies relevant for charge pumps.
An extension of the local Chern marker to interacting systems would be desirable (see \cite{Anton2020} for recent work in this direction), while polarization and entanglement spectrum can be computed in the many-body case as well (see, e.g., \cite{Hayward2018}).

Another, related  future direction would be to combine the effects of disorder and interactions and to investigate the possibility of topological charge 
pumping in both a disordered ergodic and in the many-body localized phase \cite{Abanin2019,Nandkishore2015} (see also the discussion in \cite{Wauters2019,Nakajima2020}).
Finally, the stability of topological pumping in quasiperiodic potentials other than Aubry-Andr\'{e} type might be interesting as well.

\label{sec:summary}

We acknowledge helpful discussions with M. Aidelsburger, J. Bardarson, F. Pollmann, and H. Schomerus.
We thank M. Aidelsburger for pointing out Ref.~\cite{Khemani2015} to us, J. Bardarson for bringing Refs.~\cite{Titum2016,Altland2015} to our attention,
and H. Schomerus for directing us to literature on Lifschitz tails \cite{Mezincescu1993}.
This research was funded by the Deutsche Forschungsgemeinschaft
(DFG, German Research Foundation) via Research Unit FOR 2414
under project number 277974659. U.S. acknowledges funding from EP- SRC via grant (No. EP/R044627/1) and Programme Grant DesOEQ (No. EP/P009565/1).

\appendix
\section{Additional results}
\label{sec:appendix}

\begin{figure}[t]
  \centering \includegraphics{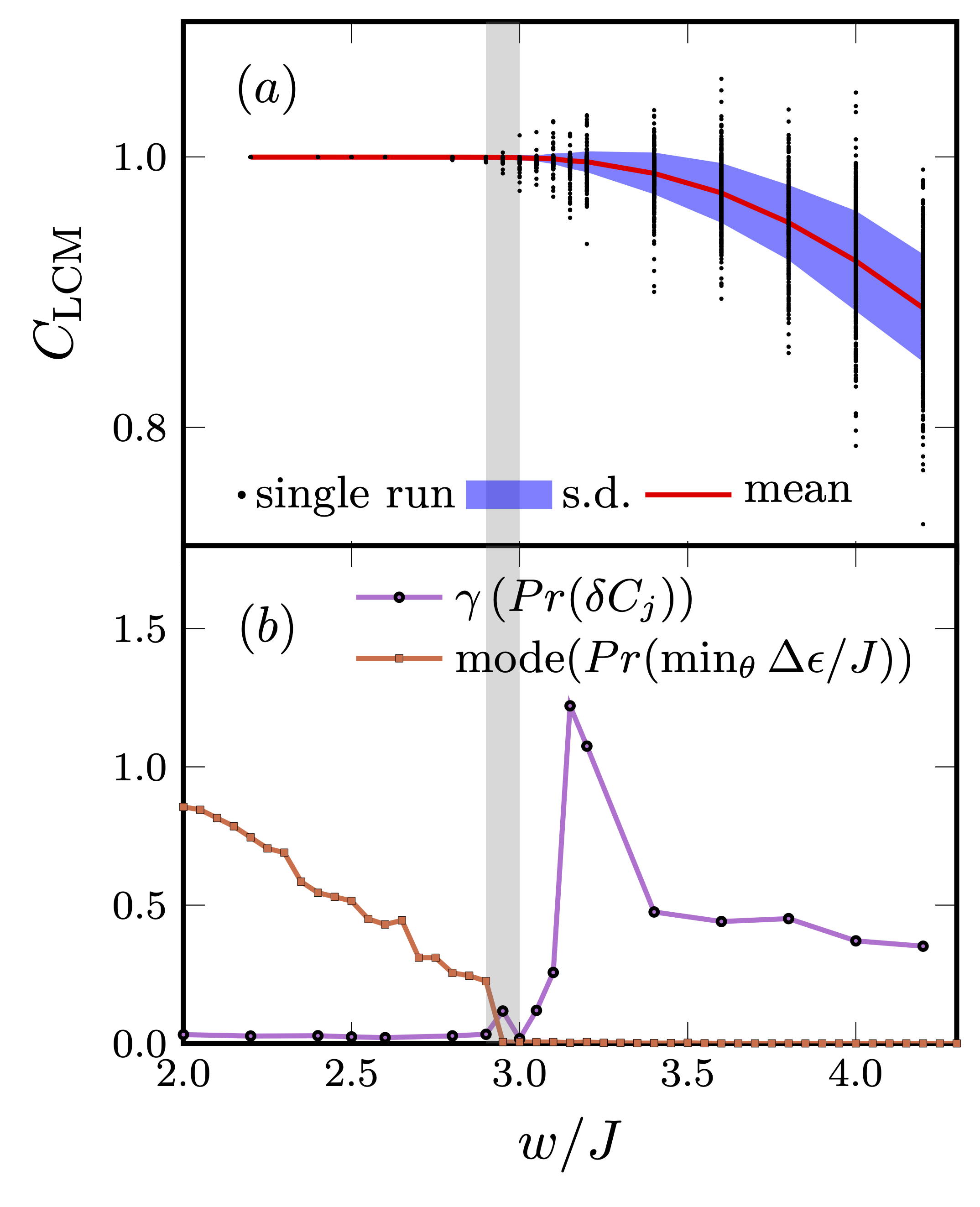}

  \mycaption{Comparison of different measures for the pumped charge for an alternative parameter set.} {For $(R_\Delta=3, R_\delta = 0.5)$, which is equivalent to parameters used in \cite{Wauters2019}, we compare (a) $\clcm$ as a function of disorder strength, (b) the skewness of LCM distributions and the mode of the minimum gap along the pump cycle. The shaded region indicates the point where the mode of the minimum gap along the pump cycle closes at $w/J\approx2.95\pm0.05$. Compared to the results of the main text, the mean of the LCM decreases slower past the transition point at $w/J\approx2.95\pm0.05$ but the transition point is unchanged. Computed for $n=500$ samples in the average for each value of $w/J$, $L=400$.
 } \label{fig:diag-average-time2}
\end{figure}

Figure~\ref{fig:diag-average-time2} shows the LCM, the skewness of the local Chern marker distributions $\gamma$, and the mode of the distribution of the minimum energy gap along the pump cycle for the parameters ($R_\Delta =3, R_\delta=0.5$), which corresponds to the parameters used in \cite{Wauters2019}. All three quantities predict a breakdown of quantized charge pumping at $w/J=2.95\pm0.05$, similar to the parameters in the main text. A possible reason for this is that $R_\delta =0.5 < R_\Delta/J$ for both parameter sets. The minimum gap during a pump cycle is therefore only controlled by $R_\delta$. Compared to the results presented in the main text, the mean of the LCM shows a shallower decrease beyond the critical disorder strength.

\newpage
\bibliographystyle{new-apsrev}
\bibliography{References}

\begin{thebibliography}{75}%
\makeatletter
\providecommand \@ifxundefined [1]{%
 \@ifx{#1\undefined}
}%
\providecommand \@ifnum [1]{%
 \ifnum #1\expandafter \@firstoftwo
 \else \expandafter \@secondoftwo
 \fi
}%
\providecommand \@ifx [1]{%
 \ifx #1\expandafter \@firstoftwo
 \else \expandafter \@secondoftwo
 \fi
}%
\providecommand \natexlab [1]{#1}%
\providecommand \enquote  [1]{``#1''}%
\providecommand \bibnamefont  [1]{#1}%
\providecommand \bibfnamefont [1]{#1}%
\providecommand \citenamefont [1]{#1}%
\providecommand \href@noop [0]{\@secondoftwo}%
\providecommand \href [0]{\begingroup \@sanitize@url \@href}%
\providecommand \@href[1]{\@@startlink{#1}\@@href}%
\providecommand \@@href[1]{\endgroup#1\@@endlink}%
\providecommand \@sanitize@url [0]{\catcode `\\12\catcode `\$12\catcode
  `\&12\catcode `\#12\catcode `\^12\catcode `\_12\catcode `\%12\relax}%
\providecommand \@@startlink[1]{}%
\providecommand \@@endlink[0]{}%
\providecommand \url  [0]{\begingroup\@sanitize@url \@url }%
\providecommand \@url [1]{\endgroup\@href {#1}{\urlprefix }}%
\providecommand \urlprefix  [0]{URL }%
\providecommand \Eprint [0]{\href }%
\providecommand \doibase [0]{http://dx.doi.org/}%
\providecommand \selectlanguage [0]{\@gobble}%
\providecommand \bibinfo  [0]{\@secondoftwo}%
\providecommand \bibfield  [0]{\@secondoftwo}%
\providecommand \translation [1]{[#1]}%
\providecommand \BibitemOpen [0]{}%
\providecommand \bibitemStop [0]{}%
\providecommand \bibitemNoStop [0]{.\EOS\space}%
\providecommand \EOS [0]{\spacefactor3000\relax}%
\providecommand \BibitemShut  [1]{\csname bibitem#1\endcsname}%
\let\auto@bib@innerbib\@empty
\bibitem [{\citenamefont {Thouless}(1983)}]{thouless1983quantization}%
  \BibitemOpen
  \bibfield  {author} {\bibinfo {author} {\bibfnamefont {D.}~\bibnamefont
  {Thouless}},\ }\href {\doibase https://doi.org/10.1103/PhysRevB.27.6083}
  {\bibfield  {journal} {\bibinfo  {journal} {Phys. Rev. B}\ }\textbf {\bibinfo
  {volume} {27}},\ \bibinfo {pages} {6083} (\bibinfo {year}
  {1983})}\BibitemShut {NoStop}%
\bibitem [{\citenamefont {Niu}\ and\ \citenamefont
  {Thouless}(1984)}]{niu1984quantised}%
  \BibitemOpen
  \bibfield  {author} {\bibinfo {author} {\bibfnamefont {Q.}~\bibnamefont
  {Niu}}\ and\ \bibinfo {author} {\bibfnamefont {D.}~\bibnamefont {Thouless}},\
  }\href {\doibase https://doi.org/10.1088/0305-4470/17/12/016} {\bibfield
  {journal} {\bibinfo  {journal} {J. Phys. A: Math. Gen.}\ }\textbf {\bibinfo
  {volume} {17}},\ \bibinfo {pages} {2453} (\bibinfo {year}
  {1984})}\BibitemShut {NoStop}%
\bibitem [{\citenamefont {Nakajima}\ \emph {et~al.}(2016)\citenamefont
  {Nakajima}, \citenamefont {Tomita}, \citenamefont {Taie}, \citenamefont
  {Ichinose}, \citenamefont {Ozawa}, \citenamefont {Wang}, \citenamefont
  {Troyer},\ and\ \citenamefont {Takahashi}}]{Nakajima2016}%
  \BibitemOpen
  \bibfield  {author} {\bibinfo {author} {\bibfnamefont {S.}~\bibnamefont
  {Nakajima}}, \bibinfo {author} {\bibfnamefont {T.}~\bibnamefont {Tomita}},
  \bibinfo {author} {\bibfnamefont {S.}~\bibnamefont {Taie}}, \bibinfo {author}
  {\bibfnamefont {T.}~\bibnamefont {Ichinose}}, \bibinfo {author}
  {\bibfnamefont {H.}~\bibnamefont {Ozawa}}, \bibinfo {author} {\bibfnamefont
  {L.}~\bibnamefont {Wang}}, \bibinfo {author} {\bibfnamefont {M.}~\bibnamefont
  {Troyer}}, \ and\ \bibinfo {author} {\bibfnamefont {Y.}~\bibnamefont
  {Takahashi}},\ }\href {\doibase https://doi.org/10.1038/nphys3622} {\bibfield
   {journal} {\bibinfo  {journal} {Nat. Phys.}\ }\textbf {\bibinfo {volume}
  {12}},\ \bibinfo {pages} {296} (\bibinfo {year} {2016})}\BibitemShut
  {NoStop}%
\bibitem [{\citenamefont {Lohse}\ \emph {et~al.}(2016)\citenamefont {Lohse},
  \citenamefont {Schweizer}, \citenamefont {Zilberberg}, \citenamefont
  {Aidelsburger},\ and\ \citenamefont {Bloch}}]{Lohse2016}%
  \BibitemOpen
  \bibfield  {author} {\bibinfo {author} {\bibfnamefont {M.}~\bibnamefont
  {Lohse}}, \bibinfo {author} {\bibfnamefont {C.}~\bibnamefont {Schweizer}},
  \bibinfo {author} {\bibfnamefont {O.}~\bibnamefont {Zilberberg}}, \bibinfo
  {author} {\bibfnamefont {M.}~\bibnamefont {Aidelsburger}}, \ and\ \bibinfo
  {author} {\bibfnamefont {I.}~\bibnamefont {Bloch}},\ }\href {\doibase
  https://doi.org/10.1038/nphys3584} {\bibfield  {journal} {\bibinfo  {journal}
  {Nat. Phys.}\ }\textbf {\bibinfo {volume} {12}},\ \bibinfo {pages} {350}
  (\bibinfo {year} {2016})}\BibitemShut {NoStop}%
\bibitem [{\citenamefont {Schweizer}\ \emph {et~al.}(2016)\citenamefont
  {Schweizer}, \citenamefont {Lohse}, \citenamefont {Citro},\ and\
  \citenamefont {Bloch}}]{Schweizer2016}%
  \BibitemOpen
  \bibfield  {author} {\bibinfo {author} {\bibfnamefont {C.}~\bibnamefont
  {Schweizer}}, \bibinfo {author} {\bibfnamefont {M.}~\bibnamefont {Lohse}},
  \bibinfo {author} {\bibfnamefont {R.}~\bibnamefont {Citro}}, \ and\ \bibinfo
  {author} {\bibfnamefont {I.}~\bibnamefont {Bloch}},\ }\href {\doibase
  10.1103/PhysRevLett.117.170405} {\bibfield  {journal} {\bibinfo  {journal}
  {Phys. Rev. Lett.}\ }\textbf {\bibinfo {volume} {117}},\ \bibinfo {pages}
  {170405} (\bibinfo {year} {2016})}\BibitemShut {NoStop}%
\bibitem [{\citenamefont {Berg}\ \emph {et~al.}(2011)\citenamefont {Berg},
  \citenamefont {Levin},\ and\ \citenamefont {Altman}}]{Berg2011}%
  \BibitemOpen
  \bibfield  {author} {\bibinfo {author} {\bibfnamefont {E.}~\bibnamefont
  {Berg}}, \bibinfo {author} {\bibfnamefont {M.}~\bibnamefont {Levin}}, \ and\
  \bibinfo {author} {\bibfnamefont {E.}~\bibnamefont {Altman}},\ }\href
  {\doibase https://doi.org/10.1103/PhysRevLett.106.110405} {\bibfield
  {journal} {\bibinfo  {journal} {Phys. Rev. Lett.}\ }\textbf {\bibinfo
  {volume} {106}},\ \bibinfo {pages} {110405} (\bibinfo {year}
  {2011})}\BibitemShut {NoStop}%
\bibitem [{\citenamefont {Rossini}\ \emph {et~al.}(2013)\citenamefont
  {Rossini}, \citenamefont {Gibertini}, \citenamefont {Giovannetti},\ and\
  \citenamefont {Fazio}}]{Rossini2013}%
  \BibitemOpen
  \bibfield  {author} {\bibinfo {author} {\bibfnamefont {D.}~\bibnamefont
  {Rossini}}, \bibinfo {author} {\bibfnamefont {M.}~\bibnamefont {Gibertini}},
  \bibinfo {author} {\bibfnamefont {V.}~\bibnamefont {Giovannetti}}, \ and\
  \bibinfo {author} {\bibfnamefont {R.}~\bibnamefont {Fazio}},\ }\href
  {\doibase 10.1103/PhysRevB.87.085131} {\bibfield  {journal} {\bibinfo
  {journal} {Phys. Rev. B}\ }\textbf {\bibinfo {volume} {87}},\ \bibinfo
  {pages} {085131} (\bibinfo {year} {2013})}\BibitemShut {NoStop}%
\bibitem [{\citenamefont {Ke}\ \emph {et~al.}(2017)\citenamefont {Ke},
  \citenamefont {Qin}, \citenamefont {Kivshar},\ and\ \citenamefont
  {Lee}}]{Ke2017}%
  \BibitemOpen
  \bibfield  {author} {\bibinfo {author} {\bibfnamefont {Y.}~\bibnamefont
  {Ke}}, \bibinfo {author} {\bibfnamefont {X.}~\bibnamefont {Qin}}, \bibinfo
  {author} {\bibfnamefont {Y.~S.}\ \bibnamefont {Kivshar}}, \ and\ \bibinfo
  {author} {\bibfnamefont {C.}~\bibnamefont {Lee}},\ }\href {\doibase
  10.1103/PhysRevA.95.063630} {\bibfield  {journal} {\bibinfo  {journal} {Phys.
  Rev. A}\ }\textbf {\bibinfo {volume} {95}},\ \bibinfo {pages} {063630}
  (\bibinfo {year} {2017})}\BibitemShut {NoStop}%
\bibitem [{\citenamefont {Kuno}\ \emph {et~al.}(2017)\citenamefont {Kuno},
  \citenamefont {Shimizu},\ and\ \citenamefont {Ichinose}}]{Kuno2017}%
  \BibitemOpen
  \bibfield  {author} {\bibinfo {author} {\bibfnamefont {Y.}~\bibnamefont
  {Kuno}}, \bibinfo {author} {\bibfnamefont {K.}~\bibnamefont {Shimizu}}, \
  and\ \bibinfo {author} {\bibfnamefont {I.}~\bibnamefont {Ichinose}},\ }\href
  {\doibase https://doi.org/10.1088/1367-2630/aa99d0} {\bibfield  {journal}
  {\bibinfo  {journal} {New J. Phys.}\ }\textbf {\bibinfo {volume} {19}},\
  \bibinfo {pages} {123025} (\bibinfo {year} {2017})}\BibitemShut {NoStop}%
\bibitem [{\citenamefont {Nakagawa}\ \emph {et~al.}(2018)\citenamefont
  {Nakagawa}, \citenamefont {Yoshida}, \citenamefont {Peters},\ and\
  \citenamefont {Kawakami}}]{Nakagawa2018}%
  \BibitemOpen
  \bibfield  {author} {\bibinfo {author} {\bibfnamefont {M.}~\bibnamefont
  {Nakagawa}}, \bibinfo {author} {\bibfnamefont {T.}~\bibnamefont {Yoshida}},
  \bibinfo {author} {\bibfnamefont {R.}~\bibnamefont {Peters}}, \ and\ \bibinfo
  {author} {\bibfnamefont {N.}~\bibnamefont {Kawakami}},\ }\href {\doibase
  10.1103/PhysRevB.98.115147} {\bibfield  {journal} {\bibinfo  {journal} {Phys.
  Rev. B}\ }\textbf {\bibinfo {volume} {98}},\ \bibinfo {pages} {115147}
  (\bibinfo {year} {2018})}\BibitemShut {NoStop}%
\bibitem [{\citenamefont {Hayward}\ \emph {et~al.}(2018)\citenamefont
  {Hayward}, \citenamefont {Schweizer}, \citenamefont {Lohse}, \citenamefont
  {Aidelsburger},\ and\ \citenamefont {Heidrich-Meisner}}]{Hayward2018}%
  \BibitemOpen
  \bibfield  {author} {\bibinfo {author} {\bibfnamefont {A.}~\bibnamefont
  {Hayward}}, \bibinfo {author} {\bibfnamefont {C.}~\bibnamefont {Schweizer}},
  \bibinfo {author} {\bibfnamefont {M.}~\bibnamefont {Lohse}}, \bibinfo
  {author} {\bibfnamefont {M.}~\bibnamefont {Aidelsburger}}, \ and\ \bibinfo
  {author} {\bibfnamefont {F.}~\bibnamefont {Heidrich-Meisner}},\ }\href
  {\doibase 10.1103/PhysRevB.98.245148} {\bibfield  {journal} {\bibinfo
  {journal} {Phys. Rev. B}\ }\textbf {\bibinfo {volume} {98}},\ \bibinfo
  {pages} {245148} (\bibinfo {year} {2018})}\BibitemShut {NoStop}%
\bibitem [{\citenamefont {Qin}\ \emph {et~al.}(2018)\citenamefont {Qin},
  \citenamefont {Schnell}, \citenamefont {Sengstock}, \citenamefont
  {Weitenberg}, \citenamefont {Eckardt},\ and\ \citenamefont
  {Hofstetter}}]{Qin2018}%
  \BibitemOpen
  \bibfield  {author} {\bibinfo {author} {\bibfnamefont {T.}~\bibnamefont
  {Qin}}, \bibinfo {author} {\bibfnamefont {A.}~\bibnamefont {Schnell}},
  \bibinfo {author} {\bibfnamefont {K.}~\bibnamefont {Sengstock}}, \bibinfo
  {author} {\bibfnamefont {C.}~\bibnamefont {Weitenberg}}, \bibinfo {author}
  {\bibfnamefont {A.}~\bibnamefont {Eckardt}}, \ and\ \bibinfo {author}
  {\bibfnamefont {W.}~\bibnamefont {Hofstetter}},\ }\href {\doibase
  10.1103/PhysRevA.98.033601} {\bibfield  {journal} {\bibinfo  {journal} {Phys.
  Rev. A}\ }\textbf {\bibinfo {volume} {98}},\ \bibinfo {pages} {033601}
  (\bibinfo {year} {2018})}\BibitemShut {NoStop}%
\bibitem [{\citenamefont {Mei}\ \emph {et~al.}(2019)\citenamefont {Mei},
  \citenamefont {Chen}, \citenamefont {Goldman}, \citenamefont {Xiao},\ and\
  \citenamefont {Jia}}]{Mei2019}%
  \BibitemOpen
  \bibfield  {author} {\bibinfo {author} {\bibfnamefont {F.}~\bibnamefont
  {Mei}}, \bibinfo {author} {\bibfnamefont {G.}~\bibnamefont {Chen}}, \bibinfo
  {author} {\bibfnamefont {N.}~\bibnamefont {Goldman}}, \bibinfo {author}
  {\bibfnamefont {L.}~\bibnamefont {Xiao}}, \ and\ \bibinfo {author}
  {\bibfnamefont {S.}~\bibnamefont {Jia}},\ }\href {\doibase
  10.1088/1367-2630/ab3d93} {\bibfield  {journal} {\bibinfo  {journal} {New J.
  Phys.}\ }\textbf {\bibinfo {volume} {21}},\ \bibinfo {pages} {095002}
  (\bibinfo {year} {2019})}\BibitemShut {NoStop}%
\bibitem [{\citenamefont {Stenzel}\ \emph {et~al.}(2019)\citenamefont
  {Stenzel}, \citenamefont {Hayward}, \citenamefont {Hubig}, \citenamefont
  {Schollw\"ock},\ and\ \citenamefont {Heidrich-Meisner}}]{Stenzel2019}%
  \BibitemOpen
  \bibfield  {author} {\bibinfo {author} {\bibfnamefont {L.}~\bibnamefont
  {Stenzel}}, \bibinfo {author} {\bibfnamefont {A.~L.~C.}\ \bibnamefont
  {Hayward}}, \bibinfo {author} {\bibfnamefont {C.}~\bibnamefont {Hubig}},
  \bibinfo {author} {\bibfnamefont {U.}~\bibnamefont {Schollw\"ock}}, \ and\
  \bibinfo {author} {\bibfnamefont {F.}~\bibnamefont {Heidrich-Meisner}},\
  }\href {\doibase 10.1103/PhysRevA.99.053614} {\bibfield  {journal} {\bibinfo
  {journal} {Phys. Rev. A}\ }\textbf {\bibinfo {volume} {99}},\ \bibinfo
  {pages} {053614} (\bibinfo {year} {2019})}\BibitemShut {NoStop}%
\bibitem [{\citenamefont {Lin}\ \emph {et~al.}(2020{\natexlab{a}})\citenamefont
  {Lin}, \citenamefont {Ke},\ and\ \citenamefont {Lee}}]{Lin2020}%
  \BibitemOpen
  \bibfield  {author} {\bibinfo {author} {\bibfnamefont {L.}~\bibnamefont
  {Lin}}, \bibinfo {author} {\bibfnamefont {Y.}~\bibnamefont {Ke}}, \ and\
  \bibinfo {author} {\bibfnamefont {C.}~\bibnamefont {Lee}},\ }\href {\doibase
  10.1103/PhysRevA.101.023620} {\bibfield  {journal} {\bibinfo  {journal}
  {Phys. Rev. A}\ }\textbf {\bibinfo {volume} {101}},\ \bibinfo {pages}
  {023620} (\bibinfo {year} {2020}{\natexlab{a}})}\BibitemShut {NoStop}%
\bibitem [{\citenamefont {Greschner}\ \emph {et~al.}(2020)\citenamefont
  {Greschner}, \citenamefont {Mondal},\ and\ \citenamefont
  {Mishra}}]{Greschner2020}%
  \BibitemOpen
  \bibfield  {author} {\bibinfo {author} {\bibfnamefont {S.}~\bibnamefont
  {Greschner}}, \bibinfo {author} {\bibfnamefont {S.}~\bibnamefont {Mondal}}, \
  and\ \bibinfo {author} {\bibfnamefont {T.}~\bibnamefont {Mishra}},\ }\href
  {\doibase 10.1103/PhysRevA.101.053630} {\bibfield  {journal} {\bibinfo
  {journal} {Phys. Rev. A}\ }\textbf {\bibinfo {volume} {101}},\ \bibinfo
  {pages} {053630} (\bibinfo {year} {2020})}\BibitemShut {NoStop}%
\bibitem [{\citenamefont {Lin}\ \emph {et~al.}(2020{\natexlab{b}})\citenamefont
  {Lin}, \citenamefont {Kennes}, \citenamefont {Pletyukhov}, \citenamefont
  {Weber}, \citenamefont {Schoeller},\ and\ \citenamefont {Meden}}]{Lin2020a}%
  \BibitemOpen
  \bibfield  {author} {\bibinfo {author} {\bibfnamefont {Y.-T.}\ \bibnamefont
  {Lin}}, \bibinfo {author} {\bibfnamefont {D.~M.}\ \bibnamefont {Kennes}},
  \bibinfo {author} {\bibfnamefont {M.}~\bibnamefont {Pletyukhov}}, \bibinfo
  {author} {\bibfnamefont {C.~S.}\ \bibnamefont {Weber}}, \bibinfo {author}
  {\bibfnamefont {H.}~\bibnamefont {Schoeller}}, \ and\ \bibinfo {author}
  {\bibfnamefont {V.}~\bibnamefont {Meden}},\ }\href {\doibase
  10.1103/PhysRevB.102.085122} {\bibfield  {journal} {\bibinfo  {journal}
  {Phys. Rev. B}\ }\textbf {\bibinfo {volume} {102}},\ \bibinfo {pages}
  {085122} (\bibinfo {year} {2020}{\natexlab{b}})}\BibitemShut {NoStop}%
\bibitem [{\citenamefont {Marks}\ \emph {et~al.}(2021)\citenamefont {Marks},
  \citenamefont {Sch\"uler}, \citenamefont {Budich},\ and\ \citenamefont
  {Devereaux}}]{Marks2021}%
  \BibitemOpen
  \bibfield  {author} {\bibinfo {author} {\bibfnamefont {J.~A.}\ \bibnamefont
  {Marks}}, \bibinfo {author} {\bibfnamefont {M.}~\bibnamefont {Sch\"uler}},
  \bibinfo {author} {\bibfnamefont {J.~C.}\ \bibnamefont {Budich}}, \ and\
  \bibinfo {author} {\bibfnamefont {T.~P.}\ \bibnamefont {Devereaux}},\ }\href
  {\doibase 10.1103/PhysRevB.103.035112} {\bibfield  {journal} {\bibinfo
  {journal} {Phys. Rev. B}\ }\textbf {\bibinfo {volume} {103}},\ \bibinfo
  {pages} {035112} (\bibinfo {year} {2021})}\BibitemShut {NoStop}%
\bibitem [{\citenamefont {Qin}\ and\ \citenamefont
  {Guo}(2016)}]{qin_quantum_2016}%
  \BibitemOpen
  \bibfield  {author} {\bibinfo {author} {\bibfnamefont {J.}~\bibnamefont
  {Qin}}\ and\ \bibinfo {author} {\bibfnamefont {H.}~\bibnamefont {Guo}},\
  }\href {\doibase 10.1016/j.physleta.2016.05.014} {\bibfield  {journal}
  {\bibinfo  {journal} {Phys. Lett. A}\ }\textbf {\bibinfo {volume} {380}},\
  \bibinfo {pages} {2317} (\bibinfo {year} {2016})}\BibitemShut {NoStop}%
\bibitem [{\citenamefont {Imura}\ \emph {et~al.}(2018)\citenamefont {Imura},
  \citenamefont {Yoshimura}, \citenamefont {Fukui},\ and\ \citenamefont
  {Hatsugai}}]{Imura2018}%
  \BibitemOpen
  \bibfield  {author} {\bibinfo {author} {\bibfnamefont {K.-I.}\ \bibnamefont
  {Imura}}, \bibinfo {author} {\bibfnamefont {Y.}~\bibnamefont {Yoshimura}},
  \bibinfo {author} {\bibfnamefont {T.}~\bibnamefont {Fukui}}, \ and\ \bibinfo
  {author} {\bibfnamefont {Y.}~\bibnamefont {Hatsugai}},\ }\href {\doibase
  10.1088/1742-6596/969/1/012133} {\bibfield  {journal} {\bibinfo  {journal}
  {J. Phys.: Conf. Series}\ }\textbf {\bibinfo {volume} {969}},\ \bibinfo
  {pages} {012133} (\bibinfo {year} {2018})}\BibitemShut {NoStop}%
\bibitem [{\citenamefont {Wang}\ and\ \citenamefont {Song}(2019)}]{Wang2019}%
  \BibitemOpen
  \bibfield  {author} {\bibinfo {author} {\bibfnamefont {R.}~\bibnamefont
  {Wang}}\ and\ \bibinfo {author} {\bibfnamefont {Z.}~\bibnamefont {Song}},\
  }\href {\doibase 10.1103/PhysRevB.100.184304} {\bibfield  {journal} {\bibinfo
   {journal} {Phys. Rev. B}\ }\textbf {\bibinfo {volume} {100}},\ \bibinfo
  {pages} {184304} (\bibinfo {year} {2019})}\BibitemShut {NoStop}%
\bibitem [{\citenamefont {Ippoliti}\ and\ \citenamefont
  {Bhatt}(2020)}]{Ippoliti2020}%
  \BibitemOpen
  \bibfield  {author} {\bibinfo {author} {\bibfnamefont {M.}~\bibnamefont
  {Ippoliti}}\ and\ \bibinfo {author} {\bibfnamefont {R.~N.}\ \bibnamefont
  {Bhatt}},\ }\href {\doibase 10.1103/PhysRevLett.124.086602} {\bibfield
  {journal} {\bibinfo  {journal} {Phys. Rev. Lett.}\ }\textbf {\bibinfo
  {volume} {124}},\ \bibinfo {pages} {086602} (\bibinfo {year}
  {2020})}\BibitemShut {NoStop}%
\bibitem [{\citenamefont {Hu}\ \emph {et~al.}(2020)\citenamefont {Hu},
  \citenamefont {Ke},\ and\ \citenamefont {Lee}}]{Hu2020}%
  \BibitemOpen
  \bibfield  {author} {\bibinfo {author} {\bibfnamefont {S.}~\bibnamefont
  {Hu}}, \bibinfo {author} {\bibfnamefont {Y.}~\bibnamefont {Ke}}, \ and\
  \bibinfo {author} {\bibfnamefont {C.}~\bibnamefont {Lee}},\ }\href {\doibase
  10.1103/PhysRevA.101.052323} {\bibfield  {journal} {\bibinfo  {journal}
  {Phys. Rev. A}\ }\textbf {\bibinfo {volume} {101}},\ \bibinfo {pages}
  {052323} (\bibinfo {year} {2020})}\BibitemShut {NoStop}%
\bibitem [{\citenamefont {Marra}\ and\ \citenamefont
  {Nitta}(2020)}]{Marra2020}%
  \BibitemOpen
  \bibfield  {author} {\bibinfo {author} {\bibfnamefont {P.}~\bibnamefont
  {Marra}}\ and\ \bibinfo {author} {\bibfnamefont {M.}~\bibnamefont {Nitta}},\
  }\href@noop {} {\bibfield  {journal} {\bibinfo  {journal} {Phys. Rev. Res.}\
  }\textbf {\bibinfo {volume} {2}} (\bibinfo {year} {2020})}\BibitemShut
  {NoStop}%
\bibitem [{\citenamefont {Arceci}\ \emph {et~al.}(2020)\citenamefont {Arceci},
  \citenamefont {Kohn}, \citenamefont {Russomanno},\ and\ \citenamefont
  {Santoro}}]{Arceci2020}%
  \BibitemOpen
  \bibfield  {author} {\bibinfo {author} {\bibfnamefont {L.}~\bibnamefont
  {Arceci}}, \bibinfo {author} {\bibfnamefont {L.}~\bibnamefont {Kohn}},
  \bibinfo {author} {\bibfnamefont {A.}~\bibnamefont {Russomanno}}, \ and\
  \bibinfo {author} {\bibfnamefont {G.~E.}\ \bibnamefont {Santoro}},\ }\href
  {\doibase 10.1088/1742-5468/ab7a25} {\bibfield  {journal} {\bibinfo
  {journal} {J. Stat. Mech.: Theory and Experiment}\ }\textbf {\bibinfo
  {volume} {2020}},\ \bibinfo {pages} {043101} (\bibinfo {year}
  {2020})}\BibitemShut {NoStop}%
\bibitem [{\citenamefont {Wawer}\ and\ \citenamefont
  {Fleischhauer}(2020)}]{Waver2020}%
  \BibitemOpen
  \bibfield  {author} {\bibinfo {author} {\bibfnamefont {L.}~\bibnamefont
  {Wawer}}\ and\ \bibinfo {author} {\bibfnamefont {M.}~\bibnamefont
  {Fleischhauer}},\ }\href {https://arxiv.org/abs/2009.04149} {\  (\bibinfo
  {year} {2020})},\ \Eprint {http://arxiv.org/abs/2009.04149}
  {arXiv:2009.04149} \BibitemShut {NoStop}%
\bibitem [{\citenamefont {Pletyukhov}\ \emph
  {et~al.}(2020{\natexlab{a}})\citenamefont {Pletyukhov}, \citenamefont
  {Kennes}, \citenamefont {Klinovaja}, \citenamefont {Loss},\ and\
  \citenamefont {Schoeller}}]{Pletyukhov2020}%
  \BibitemOpen
  \bibfield  {author} {\bibinfo {author} {\bibfnamefont {M.}~\bibnamefont
  {Pletyukhov}}, \bibinfo {author} {\bibfnamefont {D.~M.}\ \bibnamefont
  {Kennes}}, \bibinfo {author} {\bibfnamefont {J.}~\bibnamefont {Klinovaja}},
  \bibinfo {author} {\bibfnamefont {D.}~\bibnamefont {Loss}}, \ and\ \bibinfo
  {author} {\bibfnamefont {H.}~\bibnamefont {Schoeller}},\ }\href {\doibase
  10.1103/PhysRevB.101.161106} {\bibfield  {journal} {\bibinfo  {journal}
  {Phys. Rev. B}\ }\textbf {\bibinfo {volume} {101}},\ \bibinfo {pages}
  {161106} (\bibinfo {year} {2020}{\natexlab{a}})}\BibitemShut {NoStop}%
\bibitem [{\citenamefont {Pletyukhov}\ \emph
  {et~al.}(2020{\natexlab{b}})\citenamefont {Pletyukhov}, \citenamefont
  {Kennes}, \citenamefont {Klinovaja}, \citenamefont {Loss},\ and\
  \citenamefont {Schoeller}}]{Pletyukhov2020a}%
  \BibitemOpen
  \bibfield  {author} {\bibinfo {author} {\bibfnamefont {M.}~\bibnamefont
  {Pletyukhov}}, \bibinfo {author} {\bibfnamefont {D.~M.}\ \bibnamefont
  {Kennes}}, \bibinfo {author} {\bibfnamefont {J.}~\bibnamefont {Klinovaja}},
  \bibinfo {author} {\bibfnamefont {D.}~\bibnamefont {Loss}}, \ and\ \bibinfo
  {author} {\bibfnamefont {H.}~\bibnamefont {Schoeller}},\ }\href {\doibase
  10.1103/PhysRevB.101.165304} {\bibfield  {journal} {\bibinfo  {journal}
  {Phys. Rev. B}\ }\textbf {\bibinfo {volume} {101}},\ \bibinfo {pages}
  {165304} (\bibinfo {year} {2020}{\natexlab{b}})}\BibitemShut {NoStop}%
\bibitem [{\citenamefont {Weber}\ \emph {et~al.}(2021)\citenamefont {Weber},
  \citenamefont {Piasotski}, \citenamefont {Pletyukhov}, \citenamefont
  {Klinovaja}, \citenamefont {Loss}, \citenamefont {Schoeller},\ and\
  \citenamefont {Kennes}}]{Weber2021}%
  \BibitemOpen
  \bibfield  {author} {\bibinfo {author} {\bibfnamefont {C.~S.}\ \bibnamefont
  {Weber}}, \bibinfo {author} {\bibfnamefont {K.}~\bibnamefont {Piasotski}},
  \bibinfo {author} {\bibfnamefont {M.}~\bibnamefont {Pletyukhov}}, \bibinfo
  {author} {\bibfnamefont {J.}~\bibnamefont {Klinovaja}}, \bibinfo {author}
  {\bibfnamefont {D.}~\bibnamefont {Loss}}, \bibinfo {author} {\bibfnamefont
  {H.}~\bibnamefont {Schoeller}}, \ and\ \bibinfo {author} {\bibfnamefont
  {D.~M.}\ \bibnamefont {Kennes}},\ }\href {\doibase
  10.1103/PhysRevLett.126.016803} {\bibfield  {journal} {\bibinfo  {journal}
  {Phys. Rev. Lett.}\ }\textbf {\bibinfo {volume} {126}},\ \bibinfo {pages}
  {016803} (\bibinfo {year} {2021})}\BibitemShut {NoStop}%
\bibitem [{\citenamefont {Qi}\ and\ \citenamefont {Zhang}(2011)}]{Qi2011}%
  \BibitemOpen
  \bibfield  {author} {\bibinfo {author} {\bibfnamefont {X.-L.}\ \bibnamefont
  {Qi}}\ and\ \bibinfo {author} {\bibfnamefont {S.-C.}\ \bibnamefont {Zhang}},\
  }\href {\doibase 10.1103/RevModPhys.83.1057} {\bibfield  {journal} {\bibinfo
  {journal} {Rev. Mod. Phys.}\ }\textbf {\bibinfo {volume} {83}},\ \bibinfo
  {pages} {1057} (\bibinfo {year} {2011})}\BibitemShut {NoStop}%
\bibitem [{\citenamefont {Hasan}\ and\ \citenamefont {Kane}(2010)}]{Hasan2010}%
  \BibitemOpen
  \bibfield  {author} {\bibinfo {author} {\bibfnamefont {M.~Z.}\ \bibnamefont
  {Hasan}}\ and\ \bibinfo {author} {\bibfnamefont {C.~L.}\ \bibnamefont
  {Kane}},\ }\href {\doibase 10.1103/RevModPhys.82.3045} {\bibfield  {journal}
  {\bibinfo  {journal} {Rev. Mod. Phys.}\ }\textbf {\bibinfo {volume} {82}},\
  \bibinfo {pages} {3045} (\bibinfo {year} {2010})}\BibitemShut {NoStop}%
\bibitem [{\citenamefont {Chiu}\ \emph {et~al.}(2016)\citenamefont {Chiu},
  \citenamefont {Teo}, \citenamefont {Schnyder},\ and\ \citenamefont
  {Ryu}}]{Chiu2016}%
  \BibitemOpen
  \bibfield  {author} {\bibinfo {author} {\bibfnamefont {C.-K.}\ \bibnamefont
  {Chiu}}, \bibinfo {author} {\bibfnamefont {J.~C.~Y.}\ \bibnamefont {Teo}},
  \bibinfo {author} {\bibfnamefont {A.~P.}\ \bibnamefont {Schnyder}}, \ and\
  \bibinfo {author} {\bibfnamefont {S.}~\bibnamefont {Ryu}},\ }\href {\doibase
  10.1103/RevModPhys.88.035005} {\bibfield  {journal} {\bibinfo  {journal}
  {Rev. Mod. Phys.}\ }\textbf {\bibinfo {volume} {88}},\ \bibinfo {pages}
  {035005} (\bibinfo {year} {2016})}\BibitemShut {NoStop}%
\bibitem [{\citenamefont {Cooper}\ \emph {et~al.}(2019)\citenamefont {Cooper},
  \citenamefont {Dalibard},\ and\ \citenamefont {Spielman}}]{Cooper2019}%
  \BibitemOpen
  \bibfield  {author} {\bibinfo {author} {\bibfnamefont {N.~R.}\ \bibnamefont
  {Cooper}}, \bibinfo {author} {\bibfnamefont {J.}~\bibnamefont {Dalibard}}, \
  and\ \bibinfo {author} {\bibfnamefont {I.~B.}\ \bibnamefont {Spielman}},\
  }\href {\doibase 10.1103/RevModPhys.91.015005} {\bibfield  {journal}
  {\bibinfo  {journal} {Rev. Mod. Phys.}\ }\textbf {\bibinfo {volume} {91}},\
  \bibinfo {pages} {015005} (\bibinfo {year} {2019})}\BibitemShut {NoStop}%
\bibitem [{\citenamefont {Li}\ \emph {et~al.}(2009)\citenamefont {Li},
  \citenamefont {Chu}, \citenamefont {Jain},\ and\ \citenamefont
  {Shen}}]{Li2009}%
  \BibitemOpen
  \bibfield  {author} {\bibinfo {author} {\bibfnamefont {J.}~\bibnamefont
  {Li}}, \bibinfo {author} {\bibfnamefont {R.-L.}\ \bibnamefont {Chu}},
  \bibinfo {author} {\bibfnamefont {J.~K.}\ \bibnamefont {Jain}}, \ and\
  \bibinfo {author} {\bibfnamefont {S.-Q.}\ \bibnamefont {Shen}},\ }\href
  {\doibase 10.1103/PhysRevLett.102.136806} {\bibfield  {journal} {\bibinfo
  {journal} {Phys. Rev. Lett.}\ }\textbf {\bibinfo {volume} {102}},\ \bibinfo
  {pages} {136806} (\bibinfo {year} {2009})}\BibitemShut {NoStop}%
\bibitem [{\citenamefont {Meier}\ \emph {et~al.}(2018)\citenamefont {Meier},
  \citenamefont {An}, \citenamefont {Dauphin}, \citenamefont {Maffei},
  \citenamefont {Massignan}, \citenamefont {Hughes},\ and\ \citenamefont
  {Gadway}}]{Meier2018}%
  \BibitemOpen
  \bibfield  {author} {\bibinfo {author} {\bibfnamefont {E.~J.}\ \bibnamefont
  {Meier}}, \bibinfo {author} {\bibfnamefont {F.~A.}\ \bibnamefont {An}},
  \bibinfo {author} {\bibfnamefont {A.}~\bibnamefont {Dauphin}}, \bibinfo
  {author} {\bibfnamefont {M.}~\bibnamefont {Maffei}}, \bibinfo {author}
  {\bibfnamefont {P.}~\bibnamefont {Massignan}}, \bibinfo {author}
  {\bibfnamefont {T.~L.}\ \bibnamefont {Hughes}}, \ and\ \bibinfo {author}
  {\bibfnamefont {B.}~\bibnamefont {Gadway}},\ }\href {\doibase
  10.1126/science.aat3406} {\bibfield  {journal} {\bibinfo  {journal}
  {Science}\ }\textbf {\bibinfo {volume} {362}},\ \bibinfo {pages} {929}
  (\bibinfo {year} {2018})}\BibitemShut {NoStop}%
\bibitem [{\citenamefont {Nakajima}\ \emph {et~al.}(2020)\citenamefont
  {Nakajima}, \citenamefont {Takei}, \citenamefont {Sakuma}, \citenamefont
  {Kuno}, \citenamefont {Marra},\ and\ \citenamefont
  {Takahashi}}]{Nakajima2020}%
  \BibitemOpen
  \bibfield  {author} {\bibinfo {author} {\bibfnamefont {S.}~\bibnamefont
  {Nakajima}}, \bibinfo {author} {\bibfnamefont {N.}~\bibnamefont {Takei}},
  \bibinfo {author} {\bibfnamefont {K.}~\bibnamefont {Sakuma}}, \bibinfo
  {author} {\bibfnamefont {Y.}~\bibnamefont {Kuno}}, \bibinfo {author}
  {\bibfnamefont {P.}~\bibnamefont {Marra}}, \ and\ \bibinfo {author}
  {\bibfnamefont {Y.}~\bibnamefont {Takahashi}},\ }\href
  {https://arxiv.org/abs/2007.06817} {\  (\bibinfo {year} {2020})},\ \Eprint
  {http://arxiv.org/abs/2007.06817} {arXiv:2007.06817} \BibitemShut {NoStop}%
\bibitem [{\citenamefont {Titum}\ \emph {et~al.}(2016)\citenamefont {Titum},
  \citenamefont {Berg}, \citenamefont {Rudner}, \citenamefont {Refael},\ and\
  \citenamefont {Lindner}}]{Titum2016}%
  \BibitemOpen
  \bibfield  {author} {\bibinfo {author} {\bibfnamefont {P.}~\bibnamefont
  {Titum}}, \bibinfo {author} {\bibfnamefont {E.}~\bibnamefont {Berg}},
  \bibinfo {author} {\bibfnamefont {M.~S.}\ \bibnamefont {Rudner}}, \bibinfo
  {author} {\bibfnamefont {G.}~\bibnamefont {Refael}}, \ and\ \bibinfo {author}
  {\bibfnamefont {N.~H.}\ \bibnamefont {Lindner}},\ }\href {\doibase
  10.1103/PhysRevX.6.021013} {\bibfield  {journal} {\bibinfo  {journal} {Phys.
  Rev. X}\ }\textbf {\bibinfo {volume} {6}},\ \bibinfo {pages} {021013}
  (\bibinfo {year} {2016})}\BibitemShut {NoStop}%
\bibitem [{\citenamefont {Kraus}\ \emph {et~al.}(2012)\citenamefont {Kraus},
  \citenamefont {Lahini}, \citenamefont {Ringel}, \citenamefont {Verbin},\ and\
  \citenamefont {Zilberberg}}]{Kraus2012}%
  \BibitemOpen
  \bibfield  {author} {\bibinfo {author} {\bibfnamefont {Y.~E.}\ \bibnamefont
  {Kraus}}, \bibinfo {author} {\bibfnamefont {Y.}~\bibnamefont {Lahini}},
  \bibinfo {author} {\bibfnamefont {Z.}~\bibnamefont {Ringel}}, \bibinfo
  {author} {\bibfnamefont {M.}~\bibnamefont {Verbin}}, \ and\ \bibinfo {author}
  {\bibfnamefont {O.}~\bibnamefont {Zilberberg}},\ }\href {\doibase
  https://doi.org/10.1103/PhysRevLett.109.106402} {\bibfield  {journal}
  {\bibinfo  {journal} {Phys. Rev. Lett.}\ }\textbf {\bibinfo {volume} {109}},\
  \bibinfo {pages} {106402} (\bibinfo {year} {2012})}\BibitemShut {NoStop}%
\bibitem [{\citenamefont {Cerjan}\ \emph {et~al.}(2020)\citenamefont {Cerjan},
  \citenamefont {Wang}, \citenamefont {Huang}, \citenamefont {Chen},\ and\
  \citenamefont {Rechtsman}}]{Cerjan2020}%
  \BibitemOpen
  \bibfield  {author} {\bibinfo {author} {\bibfnamefont {A.}~\bibnamefont
  {Cerjan}}, \bibinfo {author} {\bibfnamefont {M.}~\bibnamefont {Wang}},
  \bibinfo {author} {\bibfnamefont {S.}~\bibnamefont {Huang}}, \bibinfo
  {author} {\bibfnamefont {K.~P.}\ \bibnamefont {Chen}}, \ and\ \bibinfo
  {author} {\bibfnamefont {M.~C.}\ \bibnamefont {Rechtsman}},\ }\href {\doibase
  10.1038/s41377-020-00408-2} {\bibfield  {journal} {\bibinfo  {journal}
  {Light: Science \& Applications}\ }\textbf {\bibinfo {volume} {9}},\ \bibinfo
  {pages} {178} (\bibinfo {year} {2020})}\BibitemShut {NoStop}%
\bibitem [{\citenamefont {Aubry}\ and\ \citenamefont
  {Andr{\'e}}(1980)}]{aubry1980analyticity}%
  \BibitemOpen
  \bibfield  {author} {\bibinfo {author} {\bibfnamefont {S.}~\bibnamefont
  {Aubry}}\ and\ \bibinfo {author} {\bibfnamefont {G.}~\bibnamefont
  {Andr{\'e}}},\ }\href
  {https://www.researchgate.net/publication/265502988_Analyticity_breaking_and_Anderson_localization_in_incommensurate_lattices}
  {\bibfield  {journal} {\bibinfo  {journal} {Ann. Israel Phys. Soc}\ }\textbf
  {\bibinfo {volume} {3}},\ \bibinfo {pages} {18} (\bibinfo {year}
  {1980})}\BibitemShut {NoStop}%
\bibitem [{\citenamefont {Schreiber}\ \emph {et~al.}(2015)\citenamefont
  {Schreiber}, \citenamefont {Hodgman}, \citenamefont {Bordia}, \citenamefont
  {L\"uschen}, \citenamefont {Fischer}, \citenamefont {Vosk}, \citenamefont
  {Altman}, \citenamefont {Schneider},\ and\ \citenamefont
  {Bloch}}]{Schreiber2015}%
  \BibitemOpen
  \bibfield  {author} {\bibinfo {author} {\bibfnamefont {M.}~\bibnamefont
  {Schreiber}}, \bibinfo {author} {\bibfnamefont {S.~S.}\ \bibnamefont
  {Hodgman}}, \bibinfo {author} {\bibfnamefont {P.}~\bibnamefont {Bordia}},
  \bibinfo {author} {\bibfnamefont {H.~P.}\ \bibnamefont {L\"uschen}}, \bibinfo
  {author} {\bibfnamefont {M.~H.}\ \bibnamefont {Fischer}}, \bibinfo {author}
  {\bibfnamefont {R.}~\bibnamefont {Vosk}}, \bibinfo {author} {\bibfnamefont
  {E.}~\bibnamefont {Altman}}, \bibinfo {author} {\bibfnamefont
  {U.}~\bibnamefont {Schneider}}, \ and\ \bibinfo {author} {\bibfnamefont
  {I.}~\bibnamefont {Bloch}},\ }\href {\doibase 10.1126/science.aaa7432}
  {\bibfield  {journal} {\bibinfo  {journal} {Science}\ }\textbf {\bibinfo
  {volume} {349}},\ \bibinfo {pages} {842} (\bibinfo {year}
  {2015})}\BibitemShut {NoStop}%
\bibitem [{\citenamefont {Rispoli}\ \emph {et~al.}(2019)\citenamefont
  {Rispoli}, \citenamefont {Lukin}, \citenamefont {Schittko}, \citenamefont
  {Kim}, \citenamefont {Tai}, \citenamefont {L{\'e}onard},\ and\ \citenamefont
  {Greiner}}]{Rispoli2019}%
  \BibitemOpen
  \bibfield  {author} {\bibinfo {author} {\bibfnamefont {M.}~\bibnamefont
  {Rispoli}}, \bibinfo {author} {\bibfnamefont {A.}~\bibnamefont {Lukin}},
  \bibinfo {author} {\bibfnamefont {R.}~\bibnamefont {Schittko}}, \bibinfo
  {author} {\bibfnamefont {S.}~\bibnamefont {Kim}}, \bibinfo {author}
  {\bibfnamefont {M.~E.}\ \bibnamefont {Tai}}, \bibinfo {author} {\bibfnamefont
  {J.}~\bibnamefont {L{\'e}onard}}, \ and\ \bibinfo {author} {\bibfnamefont
  {M.}~\bibnamefont {Greiner}},\ }\href {\doibase 10.1038/s41586-019-1527-2}
  {\bibfield  {journal} {\bibinfo  {journal} {Nature}\ }\textbf {\bibinfo
  {volume} {573}},\ \bibinfo {pages} {385} (\bibinfo {year}
  {2019})}\BibitemShut {NoStop}%
\bibitem [{\citenamefont {Bordia}\ \emph {et~al.}(2017)\citenamefont {Bordia},
  \citenamefont {L\"uschen}, \citenamefont {Scherg}, \citenamefont
  {Gopalakrishnan}, \citenamefont {Knap}, \citenamefont {Schneider},\ and\
  \citenamefont {Bloch}}]{Bordia2017}%
  \BibitemOpen
  \bibfield  {author} {\bibinfo {author} {\bibfnamefont {P.}~\bibnamefont
  {Bordia}}, \bibinfo {author} {\bibfnamefont {H.}~\bibnamefont {L\"uschen}},
  \bibinfo {author} {\bibfnamefont {S.}~\bibnamefont {Scherg}}, \bibinfo
  {author} {\bibfnamefont {S.}~\bibnamefont {Gopalakrishnan}}, \bibinfo
  {author} {\bibfnamefont {M.}~\bibnamefont {Knap}}, \bibinfo {author}
  {\bibfnamefont {U.}~\bibnamefont {Schneider}}, \ and\ \bibinfo {author}
  {\bibfnamefont {I.}~\bibnamefont {Bloch}},\ }\href {\doibase
  10.1103/PhysRevX.7.041047} {\bibfield  {journal} {\bibinfo  {journal} {Phys.
  Rev. X}\ }\textbf {\bibinfo {volume} {7}},\ \bibinfo {pages} {041047}
  (\bibinfo {year} {2017})}\BibitemShut {NoStop}%
\bibitem [{\citenamefont {Roati}\ \emph {et~al.}(2008)\citenamefont {Roati},
  \citenamefont {D'Errico}, \citenamefont {Fallani}, \citenamefont {Fattori},
  \citenamefont {Fort}, \citenamefont {Zaccanti}, \citenamefont {Modugno},
  \citenamefont {Modugno},\ and\ \citenamefont {Inguscio}}]{Roati2008}%
  \BibitemOpen
  \bibfield  {author} {\bibinfo {author} {\bibfnamefont {G.}~\bibnamefont
  {Roati}}, \bibinfo {author} {\bibfnamefont {C.}~\bibnamefont {D'Errico}},
  \bibinfo {author} {\bibfnamefont {L.}~\bibnamefont {Fallani}}, \bibinfo
  {author} {\bibfnamefont {M.}~\bibnamefont {Fattori}}, \bibinfo {author}
  {\bibfnamefont {C.}~\bibnamefont {Fort}}, \bibinfo {author} {\bibfnamefont
  {M.}~\bibnamefont {Zaccanti}}, \bibinfo {author} {\bibfnamefont
  {G.}~\bibnamefont {Modugno}}, \bibinfo {author} {\bibfnamefont
  {M.}~\bibnamefont {Modugno}}, \ and\ \bibinfo {author} {\bibfnamefont
  {M.}~\bibnamefont {Inguscio}},\ }\href {\doibase
  https://doi.org/10.1038/nature07071} {\bibfield  {journal} {\bibinfo
  {journal} {Nature (London)}\ }\textbf {\bibinfo {volume} {453}},\ \bibinfo
  {pages} {895} (\bibinfo {year} {2008})}\BibitemShut {NoStop}%
\bibitem [{\citenamefont {Dareau}\ \emph {et~al.}(2017)\citenamefont {Dareau},
  \citenamefont {Levy}, \citenamefont {Aguilera}, \citenamefont {Bouganne},
  \citenamefont {Akkermans}, \citenamefont {Gerbier},\ and\ \citenamefont
  {Beugnon}}]{Dareau2017}%
  \BibitemOpen
  \bibfield  {author} {\bibinfo {author} {\bibfnamefont {A.}~\bibnamefont
  {Dareau}}, \bibinfo {author} {\bibfnamefont {E.}~\bibnamefont {Levy}},
  \bibinfo {author} {\bibfnamefont {M.~B.}\ \bibnamefont {Aguilera}}, \bibinfo
  {author} {\bibfnamefont {R.}~\bibnamefont {Bouganne}}, \bibinfo {author}
  {\bibfnamefont {E.}~\bibnamefont {Akkermans}}, \bibinfo {author}
  {\bibfnamefont {F.}~\bibnamefont {Gerbier}}, \ and\ \bibinfo {author}
  {\bibfnamefont {J.}~\bibnamefont {Beugnon}},\ }\href {\doibase
  10.1103/PhysRevLett.119.215304} {\bibfield  {journal} {\bibinfo  {journal}
  {Phys. Rev. Lett.}\ }\textbf {\bibinfo {volume} {119}},\ \bibinfo {pages}
  {215304} (\bibinfo {year} {2017})}\BibitemShut {NoStop}%
\bibitem [{\citenamefont {Viebahn}\ \emph {et~al.}(2019)\citenamefont
  {Viebahn}, \citenamefont {Sbroscia}, \citenamefont {Carter}, \citenamefont
  {Yu},\ and\ \citenamefont {Schneider}}]{Viebahn2019}%
  \BibitemOpen
  \bibfield  {author} {\bibinfo {author} {\bibfnamefont {K.}~\bibnamefont
  {Viebahn}}, \bibinfo {author} {\bibfnamefont {M.}~\bibnamefont {Sbroscia}},
  \bibinfo {author} {\bibfnamefont {E.}~\bibnamefont {Carter}}, \bibinfo
  {author} {\bibfnamefont {J.-C.}\ \bibnamefont {Yu}}, \ and\ \bibinfo {author}
  {\bibfnamefont {U.}~\bibnamefont {Schneider}},\ }\href {\doibase
  10.1103/PhysRevLett.122.110404} {\bibfield  {journal} {\bibinfo  {journal}
  {Phys. Rev. Lett.}\ }\textbf {\bibinfo {volume} {122}},\ \bibinfo {pages}
  {110404} (\bibinfo {year} {2019})}\BibitemShut {NoStop}%
\bibitem [{\citenamefont {Rajagopal}\ \emph {et~al.}(2019)\citenamefont
  {Rajagopal}, \citenamefont {Shimasaki}, \citenamefont {Dotti}, \citenamefont
  {Ra\ifmmode \check{c}\else \v{c}\fi{}i\ifmmode~\bar{u}\else \={u}\fi{}nas},
  \citenamefont {Senaratne}, \citenamefont {Anisimovas}, \citenamefont
  {Eckardt},\ and\ \citenamefont {Weld}}]{Rajagopal2019}%
  \BibitemOpen
  \bibfield  {author} {\bibinfo {author} {\bibfnamefont {S.~V.}\ \bibnamefont
  {Rajagopal}}, \bibinfo {author} {\bibfnamefont {T.}~\bibnamefont
  {Shimasaki}}, \bibinfo {author} {\bibfnamefont {P.}~\bibnamefont {Dotti}},
  \bibinfo {author} {\bibfnamefont {M.}~\bibnamefont {Ra\ifmmode \check{c}\else
  \v{c}\fi{}i\ifmmode~\bar{u}\else \={u}\fi{}nas}}, \bibinfo {author}
  {\bibfnamefont {R.}~\bibnamefont {Senaratne}}, \bibinfo {author}
  {\bibfnamefont {E.}~\bibnamefont {Anisimovas}}, \bibinfo {author}
  {\bibfnamefont {A.}~\bibnamefont {Eckardt}}, \ and\ \bibinfo {author}
  {\bibfnamefont {D.~M.}\ \bibnamefont {Weld}},\ }\href {\doibase
  10.1103/PhysRevLett.123.223201} {\bibfield  {journal} {\bibinfo  {journal}
  {Phys. Rev. Lett.}\ }\textbf {\bibinfo {volume} {123}},\ \bibinfo {pages}
  {223201} (\bibinfo {year} {2019})}\BibitemShut {NoStop}%
\bibitem [{\citenamefont {Xiao}\ \emph {et~al.}(2020)\citenamefont {Xiao},
  \citenamefont {Xie}, \citenamefont {Dong}, \citenamefont {Chen},
  \citenamefont {Yi},\ and\ \citenamefont {Yan}}]{Xiao2020}%
  \BibitemOpen
  \bibfield  {author} {\bibinfo {author} {\bibfnamefont {T.}~\bibnamefont
  {Xiao}}, \bibinfo {author} {\bibfnamefont {D.}~\bibnamefont {Xie}}, \bibinfo
  {author} {\bibfnamefont {Z.}~\bibnamefont {Dong}}, \bibinfo {author}
  {\bibfnamefont {T.}~\bibnamefont {Chen}}, \bibinfo {author} {\bibfnamefont
  {W.}~\bibnamefont {Yi}}, \ and\ \bibinfo {author} {\bibfnamefont
  {B.}~\bibnamefont {Yan}},\ }\href {https://arxiv.org/abs/2011.03666} {\
  (\bibinfo {year} {2020})},\ \Eprint {http://arxiv.org/abs/2011.03666}
  {arXiv:2011.03666} \BibitemShut {NoStop}%
\bibitem [{\citenamefont {Liu}\ and\ \citenamefont
  {Guo}(2018)}]{liu_topological_2018}%
  \BibitemOpen
  \bibfield  {author} {\bibinfo {author} {\bibfnamefont {T.}~\bibnamefont
  {Liu}}\ and\ \bibinfo {author} {\bibfnamefont {H.}~\bibnamefont {Guo}},\
  }\href {\doibase 10.1016/j.physleta.2018.09.023} {\bibfield  {journal}
  {\bibinfo  {journal} {Physics Letters A}\ }\textbf {\bibinfo {volume}
  {382}},\ \bibinfo {pages} {3287} (\bibinfo {year} {2018})}\BibitemShut
  {NoStop}%
\bibitem [{\citenamefont {Wauters}\ \emph {et~al.}(2019)\citenamefont
  {Wauters}, \citenamefont {Russomanno}, \citenamefont {Citro}, \citenamefont
  {Santoro},\ and\ \citenamefont {Privitera}}]{Wauters2019}%
  \BibitemOpen
  \bibfield  {author} {\bibinfo {author} {\bibfnamefont {M.~M.}\ \bibnamefont
  {Wauters}}, \bibinfo {author} {\bibfnamefont {A.}~\bibnamefont {Russomanno}},
  \bibinfo {author} {\bibfnamefont {R.}~\bibnamefont {Citro}}, \bibinfo
  {author} {\bibfnamefont {G.~E.}\ \bibnamefont {Santoro}}, \ and\ \bibinfo
  {author} {\bibfnamefont {L.}~\bibnamefont {Privitera}},\ }\href {\doibase
  10.1103/PhysRevLett.123.266601} {\bibfield  {journal} {\bibinfo  {journal}
  {Phys. Rev. Lett.}\ }\textbf {\bibinfo {volume} {123}},\ \bibinfo {pages}
  {266601} (\bibinfo {year} {2019})}\BibitemShut {NoStop}%
\bibitem [{\citenamefont {Abrahams}\ \emph {et~al.}(1979)\citenamefont
  {Abrahams}, \citenamefont {Anderson}, \citenamefont {Licciardello},\ and\
  \citenamefont {Ramakrishnan}}]{Abrahams1979}%
  \BibitemOpen
  \bibfield  {author} {\bibinfo {author} {\bibfnamefont {E.}~\bibnamefont
  {Abrahams}}, \bibinfo {author} {\bibfnamefont {P.~W.}\ \bibnamefont
  {Anderson}}, \bibinfo {author} {\bibfnamefont {D.~C.}\ \bibnamefont
  {Licciardello}}, \ and\ \bibinfo {author} {\bibfnamefont {T.~V.}\
  \bibnamefont {Ramakrishnan}},\ }\href {\doibase 10.1103/PhysRevLett.42.673}
  {\bibfield  {journal} {\bibinfo  {journal} {Phys. Rev. Lett.}\ }\textbf
  {\bibinfo {volume} {42}},\ \bibinfo {pages} {673} (\bibinfo {year}
  {1979})}\BibitemShut {NoStop}%
\bibitem [{\citenamefont {Resta}(1994)}]{Resta1994}%
  \BibitemOpen
  \bibfield  {author} {\bibinfo {author} {\bibfnamefont {R.}~\bibnamefont
  {Resta}},\ }\href {\doibase 10.1103/RevModPhys.66.899} {\bibfield  {journal}
  {\bibinfo  {journal} {Rev. Mod. Phys.}\ }\textbf {\bibinfo {volume} {66}},\
  \bibinfo {pages} {899} (\bibinfo {year} {1994})}\BibitemShut {NoStop}%
\bibitem [{\citenamefont {Su}\ \emph {et~al.}(1979)\citenamefont {Su},
  \citenamefont {Schrieffer},\ and\ \citenamefont {Heeger}}]{Su1979}%
  \BibitemOpen
  \bibfield  {author} {\bibinfo {author} {\bibfnamefont {W.~P.}\ \bibnamefont
  {Su}}, \bibinfo {author} {\bibfnamefont {J.~R.}\ \bibnamefont {Schrieffer}},
  \ and\ \bibinfo {author} {\bibfnamefont {A.~J.}\ \bibnamefont {Heeger}},\
  }\href {\doibase 10.1103/PhysRevLett.42.1698} {\bibfield  {journal} {\bibinfo
   {journal} {Phys. Rev. Lett.}\ }\textbf {\bibinfo {volume} {42}},\ \bibinfo
  {pages} {1698} (\bibinfo {year} {1979})}\BibitemShut {NoStop}%
\bibitem [{\citenamefont {Altland}\ \emph {et~al.}(2015)\citenamefont
  {Altland}, \citenamefont {Bagrets},\ and\ \citenamefont
  {Kamenev}}]{Altland2015}%
  \BibitemOpen
  \bibfield  {author} {\bibinfo {author} {\bibfnamefont {A.}~\bibnamefont
  {Altland}}, \bibinfo {author} {\bibfnamefont {D.}~\bibnamefont {Bagrets}}, \
  and\ \bibinfo {author} {\bibfnamefont {A.}~\bibnamefont {Kamenev}},\ }\href
  {\doibase 10.1103/PhysRevB.91.085429} {\bibfield  {journal} {\bibinfo
  {journal} {Phys. Rev. B}\ }\textbf {\bibinfo {volume} {91}},\ \bibinfo
  {pages} {085429} (\bibinfo {year} {2015})}\BibitemShut {NoStop}%
\bibitem [{\citenamefont {Bianco}\ and\ \citenamefont
  {Resta}(2011)}]{Bianco2011}%
  \BibitemOpen
  \bibfield  {author} {\bibinfo {author} {\bibfnamefont {R.}~\bibnamefont
  {Bianco}}\ and\ \bibinfo {author} {\bibfnamefont {R.}~\bibnamefont {Resta}},\
  }\href {\doibase 10.1103/PhysRevB.84.241106} {\bibfield  {journal} {\bibinfo
  {journal} {Phys. Rev. B}\ }\textbf {\bibinfo {volume} {84}},\ \bibinfo
  {pages} {241106} (\bibinfo {year} {2011})}\BibitemShut {NoStop}%
\bibitem [{\citenamefont {Irsigler}\ \emph {et~al.}(2019)\citenamefont
  {Irsigler}, \citenamefont {Zheng},\ and\ \citenamefont
  {Hofstetter}}]{Irsigler2019}%
  \BibitemOpen
  \bibfield  {author} {\bibinfo {author} {\bibfnamefont {B.}~\bibnamefont
  {Irsigler}}, \bibinfo {author} {\bibfnamefont {J.-H.}\ \bibnamefont {Zheng}},
  \ and\ \bibinfo {author} {\bibfnamefont {W.}~\bibnamefont {Hofstetter}},\
  }\href {\doibase 10.1103/PhysRevLett.122.010406} {\bibfield  {journal}
  {\bibinfo  {journal} {Phys. Rev. Lett.}\ }\textbf {\bibinfo {volume} {122}},\
  \bibinfo {pages} {010406} (\bibinfo {year} {2019})}\BibitemShut {NoStop}%
\bibitem [{\citenamefont {Sykes}\ and\ \citenamefont
  {Barnett}(2020)}]{sykes2020}%
  \BibitemOpen
  \bibfield  {author} {\bibinfo {author} {\bibfnamefont {J.}~\bibnamefont
  {Sykes}}\ and\ \bibinfo {author} {\bibfnamefont {R.}~\bibnamefont
  {Barnett}},\ }\href@noop {} {\enquote {\bibinfo {title} {Local topological
  markers in odd dimensions},}\ } (\bibinfo {year} {2020}),\ \Eprint
  {http://arxiv.org/abs/2011.04771} {arXiv:2011.04771} \BibitemShut {NoStop}%
\bibitem [{\citenamefont {Rice}\ and\ \citenamefont {Mele}(1982)}]{Rice1982}%
  \BibitemOpen
  \bibfield  {author} {\bibinfo {author} {\bibfnamefont {M.}~\bibnamefont
  {Rice}}\ and\ \bibinfo {author} {\bibfnamefont {E.}~\bibnamefont {Mele}},\
  }\href {\doibase https://doi.org/10.1103/PhysRevLett.49.1455} {\bibfield
  {journal} {\bibinfo  {journal} {Phys. Rev. Lett.}\ }\textbf {\bibinfo
  {volume} {49}},\ \bibinfo {pages} {1455} (\bibinfo {year}
  {1982})}\BibitemShut {NoStop}%
\bibitem [{\citenamefont {Resta}(1998)}]{Resta1998}%
  \BibitemOpen
  \bibfield  {author} {\bibinfo {author} {\bibfnamefont {R.}~\bibnamefont
  {Resta}},\ }\href {\doibase 10.1103/PhysRevLett.80.1800} {\bibfield
  {journal} {\bibinfo  {journal} {Phys. Rev. Lett.}\ }\textbf {\bibinfo
  {volume} {80}},\ \bibinfo {pages} {1800} (\bibinfo {year}
  {1998})}\BibitemShut {NoStop}%
\bibitem [{\citenamefont {Resta}(1992)}]{Resta1992}%
  \BibitemOpen
  \bibfield  {author} {\bibinfo {author} {\bibfnamefont {R.}~\bibnamefont
  {Resta}},\ }\href {\doibase https://doi.org/10.1080/00150199208016065}
  {\bibfield  {journal} {\bibinfo  {journal} {Ferroelectrics}\ }\textbf
  {\bibinfo {volume} {136}},\ \bibinfo {pages} {51} (\bibinfo {year}
  {1992})}\BibitemShut {NoStop}%
\bibitem [{\citenamefont {King-Smith}\ and\ \citenamefont
  {Vanderbilt}(1993)}]{King1993}%
  \BibitemOpen
  \bibfield  {author} {\bibinfo {author} {\bibfnamefont {R.}~\bibnamefont
  {King-Smith}}\ and\ \bibinfo {author} {\bibfnamefont {D.}~\bibnamefont
  {Vanderbilt}},\ }\href {\doibase https://doi.org/10.1103/PhysRevB.47.1651}
  {\bibfield  {journal} {\bibinfo  {journal} {Phys. Rev. B}\ }\textbf {\bibinfo
  {volume} {47}},\ \bibinfo {pages} {1651} (\bibinfo {year}
  {1993})}\BibitemShut {NoStop}%
\bibitem [{\citenamefont {Watanabe}\ and\ \citenamefont
  {Oshikawa}(2018)}]{watanabe_inequivalent_2018}%
  \BibitemOpen
  \bibfield  {author} {\bibinfo {author} {\bibfnamefont {H.}~\bibnamefont
  {Watanabe}}\ and\ \bibinfo {author} {\bibfnamefont {M.}~\bibnamefont
  {Oshikawa}},\ }\href {\doibase 10.1103/PhysRevX.8.021065} {\bibfield
  {journal} {\bibinfo  {journal} {Phys. Rev. X}\ }\textbf {\bibinfo {volume}
  {8}},\ \bibinfo {pages} {021065} (\bibinfo {year} {2018})}\BibitemShut
  {NoStop}%
\bibitem [{\citenamefont {Li}\ and\ \citenamefont {Haldane}(2008)}]{Li2008}%
  \BibitemOpen
  \bibfield  {author} {\bibinfo {author} {\bibfnamefont {H.}~\bibnamefont
  {Li}}\ and\ \bibinfo {author} {\bibfnamefont {F.~D.~M.}\ \bibnamefont
  {Haldane}},\ }\href {\doibase https://doi.org/10.1103/PhysRevLett.101.010504}
  {\bibfield  {journal} {\bibinfo  {journal} {Phys. Rev. Lett.}\ }\textbf
  {\bibinfo {volume} {101}},\ \bibinfo {pages} {010504} (\bibinfo {year}
  {2008})}\BibitemShut {NoStop}%
\bibitem [{\citenamefont {Peschel}\ and\ \citenamefont
  {Eisler}(2009)}]{Peschel2009}%
  \BibitemOpen
  \bibfield  {author} {\bibinfo {author} {\bibfnamefont {I.}~\bibnamefont
  {Peschel}}\ and\ \bibinfo {author} {\bibfnamefont {V.}~\bibnamefont
  {Eisler}},\ }\href {\doibase 10.1088/1751-8113/42/50/504003} {\bibfield
  {journal} {\bibinfo  {journal} {J. Phys. A: Math. Theor.}\ }\textbf {\bibinfo
  {volume} {42}},\ \bibinfo {pages} {504003} (\bibinfo {year}
  {2009})}\BibitemShut {NoStop}%
\bibitem [{\citenamefont {Privitera}\ \emph {et~al.}(2018)\citenamefont
  {Privitera}, \citenamefont {Russomanno}, \citenamefont {Citro},\ and\
  \citenamefont {Santoro}}]{Privitera2018}%
  \BibitemOpen
  \bibfield  {author} {\bibinfo {author} {\bibfnamefont {L.}~\bibnamefont
  {Privitera}}, \bibinfo {author} {\bibfnamefont {A.}~\bibnamefont
  {Russomanno}}, \bibinfo {author} {\bibfnamefont {R.}~\bibnamefont {Citro}}, \
  and\ \bibinfo {author} {\bibfnamefont {G.~E.}\ \bibnamefont {Santoro}},\
  }\href {\doibase https://doi.org/10.1103/PhysRevLett.120.106601} {\bibfield
  {journal} {\bibinfo  {journal} {Phys. Rev. Lett.}\ }\textbf {\bibinfo
  {volume} {120}},\ \bibinfo {pages} {106601} (\bibinfo {year}
  {2018})}\BibitemShut {NoStop}%
\bibitem [{\citenamefont {Kuno}(2019)}]{Kuno2019}%
  \BibitemOpen
  \bibfield  {author} {\bibinfo {author} {\bibfnamefont {Y.}~\bibnamefont
  {Kuno}},\ }\href {\doibase 10.1140/epjb/e2019-100131-1} {\bibfield  {journal}
  {\bibinfo  {journal} {Eur. Phys. J. B}\ }\textbf {\bibinfo {volume} {92}},\
  \bibinfo {pages} {195} (\bibinfo {year} {2019})}\BibitemShut {NoStop}%
\bibitem [{\citenamefont {Manmana}\ \emph {et~al.}(2005)\citenamefont
  {Manmana}, \citenamefont {Muramatsu},\ and\ \citenamefont
  {Noack}}]{Manmana2005}%
  \BibitemOpen
  \bibfield  {author} {\bibinfo {author} {\bibfnamefont {S.~R.}\ \bibnamefont
  {Manmana}}, \bibinfo {author} {\bibfnamefont {A.}~\bibnamefont {Muramatsu}},
  \ and\ \bibinfo {author} {\bibfnamefont {R.~M.}\ \bibnamefont {Noack}},\
  }\href {\doibase 10.1063/1.2080353} {\bibfield  {journal} {\bibinfo
  {journal} {AIP Conf. Proc.}\ }\textbf {\bibinfo {volume} {789}},\ \bibinfo
  {pages} {269} (\bibinfo {year} {2005})}\BibitemShut {NoStop}%
\bibitem [{\citenamefont {Anderson}(1958)}]{anderson_absence_1958}%
  \BibitemOpen
  \bibfield  {author} {\bibinfo {author} {\bibfnamefont {P.~W.}\ \bibnamefont
  {Anderson}},\ }\href {\doibase 10.1103/PhysRev.109.1492} {\bibfield
  {journal} {\bibinfo  {journal} {Phys. Rev.}\ }\textbf {\bibinfo {volume}
  {109}},\ \bibinfo {pages} {1492} (\bibinfo {year} {1958})}\BibitemShut
  {NoStop}%
\bibitem [{\citenamefont {Li}\ and\ \citenamefont
  {Fleischhauer}(2017)}]{Li2017}%
  \BibitemOpen
  \bibfield  {author} {\bibinfo {author} {\bibfnamefont {R.}~\bibnamefont
  {Li}}\ and\ \bibinfo {author} {\bibfnamefont {M.}~\bibnamefont
  {Fleischhauer}},\ }\href {\doibase 10.1103/PhysRevB.96.085444} {\bibfield
  {journal} {\bibinfo  {journal} {Phys. Rev. B}\ }\textbf {\bibinfo {volume}
  {96}},\ \bibinfo {pages} {085444} (\bibinfo {year} {2017})}\BibitemShut
  {NoStop}%
\bibitem [{\citenamefont {Khemani}\ \emph {et~al.}(2015)\citenamefont
  {Khemani}, \citenamefont {Nandkishore},\ and\ \citenamefont
  {Sondhi}}]{Khemani2015}%
  \BibitemOpen
  \bibfield  {author} {\bibinfo {author} {\bibfnamefont {V.}~\bibnamefont
  {Khemani}}, \bibinfo {author} {\bibfnamefont {R.}~\bibnamefont
  {Nandkishore}}, \ and\ \bibinfo {author} {\bibfnamefont {S.}~\bibnamefont
  {Sondhi}},\ }\href {\doibase https://doi.org/10.1038/nphys3344} {\bibfield
  {journal} {\bibinfo  {journal} {Nat. Phys.}\ }\textbf {\bibinfo {volume}
  {11}},\ \bibinfo {pages} {560} (\bibinfo {year} {2015})}\BibitemShut
  {NoStop}%
\bibitem [{\citenamefont {Mezincescu}(1993)}]{Mezincescu1993}%
  \BibitemOpen
  \bibfield  {author} {\bibinfo {author} {\bibfnamefont {G.}~\bibnamefont
  {Mezincescu}},\ }\href {https://link.springer.com/article/10.1007/BF02108077}
  {\bibfield  {journal} {\bibinfo  {journal} {Commun. Math. Phys.}\ }\textbf
  {\bibinfo {volume} {158}},\ \bibinfo {pages} {315} (\bibinfo {year}
  {1993})}\BibitemShut {NoStop}%
\bibitem [{\citenamefont {Eckardt}(2017)}]{Eckardt2017}%
  \BibitemOpen
  \bibfield  {author} {\bibinfo {author} {\bibfnamefont {A.}~\bibnamefont
  {Eckardt}},\ }\href {\doibase 10.1103/RevModPhys.89.011004} {\bibfield
  {journal} {\bibinfo  {journal} {Rev. Mod. Phys.}\ }\textbf {\bibinfo {volume}
  {89}},\ \bibinfo {pages} {011004} (\bibinfo {year} {2017})}\BibitemShut
  {NoStop}%
\bibitem [{\citenamefont {Anton}\ and\ \citenamefont
  {Rubtsov}(2020)}]{Anton2020}%
  \BibitemOpen
  \bibfield  {author} {\bibinfo {author} {\bibfnamefont {M.}~\bibnamefont
  {Anton}}\ and\ \bibinfo {author} {\bibfnamefont {A.}~\bibnamefont
  {Rubtsov}},\ }\href@noop {} {\  (\bibinfo {year} {2020})},\ \Eprint
  {http://arxiv.org/abs/2009.01801} {arXiv:2009.01801} \BibitemShut {NoStop}%
\bibitem [{\citenamefont {Abanin}\ \emph {et~al.}(2019)\citenamefont {Abanin},
  \citenamefont {Altman}, \citenamefont {Bloch},\ and\ \citenamefont
  {Serbyn}}]{Abanin2019}%
  \BibitemOpen
  \bibfield  {author} {\bibinfo {author} {\bibfnamefont {D.~A.}\ \bibnamefont
  {Abanin}}, \bibinfo {author} {\bibfnamefont {E.}~\bibnamefont {Altman}},
  \bibinfo {author} {\bibfnamefont {I.}~\bibnamefont {Bloch}}, \ and\ \bibinfo
  {author} {\bibfnamefont {M.}~\bibnamefont {Serbyn}},\ }\href {\doibase
  10.1103/RevModPhys.91.021001} {\bibfield  {journal} {\bibinfo  {journal}
  {Rev. Mod. Phys.}\ }\textbf {\bibinfo {volume} {91}},\ \bibinfo {pages}
  {021001} (\bibinfo {year} {2019})}\BibitemShut {NoStop}%
\bibitem [{\citenamefont {Nandikishore}\ and\ \citenamefont
  {Huse}(2015)}]{Nandkishore2015}%
  \BibitemOpen
  \bibfield  {author} {\bibinfo {author} {\bibfnamefont {R.}~\bibnamefont
  {Nandikishore}}\ and\ \bibinfo {author} {\bibfnamefont {D.}~\bibnamefont
  {Huse}},\ }\href {\doibase 10.1146/annurev-conmatphys-031214-014726}
  {\bibfield  {journal} {\bibinfo  {journal} {Annu. Rev. Condens. Matter
  Phys.}\ }\textbf {\bibinfo {volume} {6}},\ \bibinfo {pages} {15} (\bibinfo
  {year} {2015})}\BibitemShut {NoStop}%
\end{thebibliography}%

\end{document}